\documentclass[conference,compsoc]{IEEEtran}

\usepackage[breaklinks,colorlinks, citecolor=blue, urlcolor=black, linkcolor=blue, citecolor=blue]{hyperref}
\usepackage{color}
\usepackage{amsmath,amsopn}
\usepackage{amssymb}
\usepackage[ruled,boxed,commentsnumbered, linesnumbered]{algorithm2e}
\usepackage[labelformat=simple]{subcaption}

\usepackage{endnotes,microtype,xspace,graphicx,fancyvrb,multirow}
\usepackage{booktabs}
\usepackage{array,underscore,relsize}
\usepackage[T1]{fontenc}
\usepackage{fancyhdr,totpages}
\usepackage{enumitem}
\usepackage[font=small, labelfont={bf}]{caption}
\usepackage{tabularx}
\usepackage{colortbl}

\usepackage{tcolorbox}
\usepackage[sort]{cite}

\usepackage{fp}
\usepackage{siunitx}
\usepackage{diagbox}
\usepackage[varqu,varl]{inconsolata} %

\usepackage{listings}

\AtBeginEnvironment{listing}{\setcounter{listing}{\value{lstlisting}}} 
\AtEndEnvironment{listing}{\stepcounter{lstlisting}}

\usepackage{balance}

\definecolor{lightgray}{rgb}{0.91, 0.91, 0.91}
\definecolor{lightblue}{rgb}{0.8627, 0.9176, 0.9686} %

\usepackage{multicol}

\usepackage{verbatim}

\renewcommand{\arraystretch}{1.15}  %

\usepackage[normalem]{ulem}

\definecolor{linkblue}{RGB}{40,80,160}
\hypersetup{linkcolor=linkblue, citecolor=linkblue}

\newcommand{\sys}{\mbox{\textsc{AntiAgent}}\xspace}

\usepackage[dvipsnames,table]{xcolor}

\newcommand{\agenticon}[2][0.95em]{%
    \makebox[1.65em][c]{%
        \raisebox{0.3em}{%
            \raisebox{-0.5\height}{%
                \includegraphics[height=#1]{#2}%
            }%
        }%
    }%
    \hspace{0.3em}%
}

\newcommand{\bestdsr}[1]{\cellcolor{black!8}\textbf{#1}}

\usepackage{url}
\newcommand{\cc}[1]{\mbox{\smaller[0.5]\texttt{#1}}}

\fvset{fontsize=\scriptsize,xleftmargin=8pt,numbers=left,numbersep=5pt}

\def\Snospace~{\S{}}

\newcommand{\x}{$\times$\xspace}

\newif\ifdraft\drafttrue
\newif\ifnotes\notestrue
\ifdraft\else\notesfalse\fi

\input{glyphtounicode}
\newcommand{\squishlist}{
\begin{itemize}[noitemsep,nolistsep,leftmargin=10pt]
  \setlength{\itemsep}{-0pt}
}
\newcommand{\squishend}{
  \end{itemize}
}

\usepackage{tikz}
\newcommand*\WC[1]{%
\begin{tikzpicture}[baseline=(C.base)]
\node[draw,circle,inner sep=0.2pt](C) {#1};
\end{tikzpicture}}

\newcommand*\emptycirc[1][1ex]{\tikz\draw (0,0) circle (#1);} 
\newcommand*\halfcirc[1][1ex]{%
  \begin{tikzpicture}
  \draw[fill] (0,0)-- (90:#1) arc (90:270:#1) -- cycle ;
  \draw (0,0) circle (#1);
  \end{tikzpicture}}
\newcommand*\fullcirc[1][1.05ex]{\tikz\fill (0,0) circle (#1);}

\usepackage{xstring}
\newcommand{\PP}[1]{%
  \begingroup
    \edef\temp{\detokenize{#1}}%
    \IfEndWith{\temp}{.}{%
      \noindent\textbf{#1}%
    }{%
      \noindent\textbf{#1.}%
    }%
  \endgroup
}

\newcommand{\PN}[1]{
\noindent{\bf #1}
}

\newtcolorbox{mybox}[3][float=h]
{
  colback=#2!5!white,
  colbacktitle=#2!15!white,
  boxrule=0.25mm, 
  top=0pt, bottom=0pt, left=0pt, right=0pt,
  coltitle=black,
  title    = {#3},
  #1,
}

\newcounter{prompt}
\renewcommand{\theprompt}{\arabic{prompt}}  %

\newcommand{\promptbox}[2]{ %
\refstepcounter{prompt}
\begin{mybox}{Melon}{\textbf{\small{Prompt \theprompt: #1}}}
{#2}
\end{mybox}
}

\newcommand{\appautoref}[1]{\hyperref[#1]{Appendix~\ref{#1}}}

\begin{document}

\title{Hardening JavaScript Programs against LLM-Based Vulnerability Detection}

\newcommand{\authorrowbreak}{%
    \and[\hfill\mbox{}\par\vspace{1em}\mbox{}\hfill]%
}

\author{
\IEEEauthorblockN{Zhenhao Tian}
\IEEEauthorblockA{
Nanyang Technological University\\
Tsinghua University\\
tian-zh24@mails.tsinghua.edu.cn
}
\and
\IEEEauthorblockN{Jintong Yu}
\IEEEauthorblockA{
Nanyang Technological University\\
Shanghai Jiao Tong University\\
jintongyu@sjtu.edu.cn
}
\and
\IEEEauthorblockN{Jingwei Feng}
\IEEEauthorblockA{
Tsinghua University\\
fengjw26@mails.tsinghua.edu.cn
}

\authorrowbreak

\IEEEauthorblockN{Jianwei Zhuge}
\IEEEauthorblockA{
Tsinghua University\\
zhugejw@tsinghua.edu.cn
}
\and
\IEEEauthorblockN{Penghui Li}
\IEEEauthorblockA{
Nanyang Technological University\\
penghui.li@ntu.edu.sg
}
}

\date{}
\maketitle

\begin{abstract}
JavaScript programs are often distributed as source code, so attackers can analyze them for vulnerabilities.
LLM-based vulnerability detection agents now make this analysis much easier.
To mitigate this emerging threat, we present \sys, a defense that lets developers harden their programs against such detection agents.
\sys transforms a program before release to impede an agent's analysis.
To find effective transformations, \sys evolves a corpus of programs with known vulnerabilities and traces how a detection agent analyzes each evolved version.
It keeps the versions on which the agent misses or misreports the vulnerability and extracts their transformations into reusable rules.
To protect a new program, \sys carefully divides the task between an LLM and the rules, so that the transformations adapt to the program while preserving its semantics.

We evaluate \sys on real-world JavaScript applications against four detection agents, including Claude Code and Codex.
\sys reduces the number of vulnerabilities that each agent detects by 52.0--60.7\%, outperforming the conventional obfuscator JS-Obfuscator and the LLM-based transformation method FPA on every agent.
The transformed programs pass all project tests, indicating that \sys preserves the original program semantics.
The transformation rules also remain effective on vulnerability types absent from the corpus and on programs written in another language like Python.

\end{abstract}

\section{Introduction}
\label{s:intro}
JavaScript is widely used in web applications to implement dynamic features in browsers and server-side runtimes such as Node.js~\cite{rack2023jack,secbench:icse:2023}.
Client-side JavaScript is delivered to browsers and executes on users' devices~\cite{skolka2019anything}.
Vulnerabilities in JavaScript programs can expose sensitive data~\cite{owasp:frontend-security}, enable session hijacking~\cite{owasp:cookie-theft}, or allow unauthorized actions on a user's behalf~\cite{owasp:csrf,khodayari2021jaw}.

JavaScript is often distributed as source code, so anyone who obtains the code can analyze it for vulnerabilities~\cite{shcherbakov2023silentspring}.
Agents built on large language models (LLMs) now make this analysis much easier~\cite{repoaudit, enigma, bugauditor2026, ni2026llmvdjs, simsek2025pocgen}, and attackers have begun to use them for malicious vulnerability analysis.
OALABS researchers analyzed attack logs and found the use of Claude Code~\cite{claude-code} and Codex~\cite{codex} to identify vulnerable services and develop exploits~\cite{ai-hacker-14-companies}.
Anthropic also reports a software-as-a-service intrusion in which an attacker exploited a cross-site scripting (XSS) vulnerability, a flaw that can allow attacker-controlled JavaScript to execute in a victim's browser~\cite{anthropic-threat-report, owasp:frontend-security}.
The attacker then used Claude Code to explore the service's backend APIs and export customer data~\cite{anthropic-threat-report}.

These incidents highlight the need for defenses that protect JavaScript programs against LLM-based vulnerability detection.
Developers cannot prevent attackers from analyzing these programs with LLM agents, nor can they fix vulnerabilities they have not yet found.
Instead, they can harden their programs against such analysis before release, so that agents are less likely to detect vulnerabilities in them.
This mirrors anti-fuzzing, which impedes fuzzers on released binaries while preserving normal execution~\cite{jung2019fuzzification}.

In this paper, we explore this hardening approach through code transformations, which directly modify a program's source code.
Unlike adversarial attacks~\cite{attack_1_hidenoseek, attack_2_wolf} on code models, which modify malicious code to evade detection, such hardening modifies benign programs to protect them from the attackers.
To serve as a practical defense, these transformations should reduce the vulnerabilities that detection agents detect and, more importantly, preserve the program's semantics.
They should also add little overhead and generalize across programs, vulnerability types, and detection agents.

No existing transformation meets these requirements.
Minification and conventional obfuscators such as JS-Obfuscator change only the surface syntax, by shortening names, encoding strings, or flattening control flow~\cite{skolka2019anything,javascript-obfuscator}.
These changes follow fixed patterns, so a capable agent recognizes them and still traces the vulnerable flow.
JS-Obfuscator also more than doubles execution time and code size in its \cc{medium} configuration.
Recent methods instead target the reasoning of LLMs.
Familiar Pattern Attack (FPA)~\cite{fpa} embeds an altered familiar pattern into a branch condition, tricking the model into following the wrong execution path.
CoTDeceptor~\cite{li2025cotdeceptor} searches each program anew, using the detector's chain-of-thought to guide transformations until it misses the vulnerability.
CoTDeceptor also requires knowing the vulnerabilities in advance, which is impractical for defense.
Both also do not ensure their transformations preserve behavior.
For example, FPA's transformed programs pass only 68.52\% of project tests in our evaluation.

We present \sys, a defense that protects JavaScript programs through code transformations.
Our key insight is that effectiveness requires novel transformation rules that a detection agent cannot easily recognize and bypass.
On the other hand, semantics preservation requires that the transformations be restricted to predefined operations rather than applied freely, so that they do not change the program's original behavior.
Therefore, \sys discovers novel transformations by search over corpus programs and extracts them into reusable rules.
Each rule specifies when and how to change the code while preserving the program's behavior.
\sys then applies the rules to new programs, where an LLM dynamically selects a fitting rule to make the code changes.

We evaluate \sys on 40 real-world JavaScript applications that average about 37K lines of code and span 13 vulnerability types.
All vulnerabilities were disclosed after the release of the evaluated models to avoid data contamination.
We use four detection agents, including the general coding agents Codex~\cite{codex}, Claude Code~\cite{claude-code}, and ZCode~\cite{zcode}, and the dedicated JavaScript detector OpenAnt~\cite{openant}.
\sys reduces the number of vulnerabilities that the four agents detect by 52.0--60.7\%, more than JS-Obfuscator~\cite{javascript-obfuscator} and FPA~\cite{fpa} on every agent.
For example, Claude Code detects every vulnerability in the original programs but only 45.0\% of them after \sys is applied.
The transformed programs pass all project tests, whereas JS-Obfuscator and FPA pass only 88.46\% and 68.52\% of the tests, respectively.
\sys-transformed programs run in 1.11\x the original execution time and are 1.37\x the original size on average, compared with 2.46\x and 2.63\x for JS-Obfuscator.
We show that the rules remain effective on different vulnerability types and on programs written in another language Python.

Our contributions are summarized as follows.
\squishlist
    \item We design an evolutionary search that discovers code transformations impeding detection agents.
    \item We develop \sys, a system that turns these transformations into reusable rules and applies them to new programs while preserving semantics.
    \item We evaluate \sys on real-world JavaScript applications, where it reduces detection by 52.0--60.7\% across four detection agents.
\squishend

\section{Background}
\label{s:background}

\subsection{LLM-Based Vulnerability Detection Agents}
LLM-based vulnerability detection agents combine a language model with tools for searching and reading source files, following data flows, executing code, and running tests~\cite{repoaudit,enigma}.
The model generates hypotheses about potential vulnerabilities, chooses tool actions~\cite{shi2026toolhijacker} to test them, and refines the hypotheses based on the results~\cite{react}.
Unlike a classifier that labels a fixed code fragment, an agent investigates a whole repository and validates its findings interactively~\cite{sweagent}.
It typically outputs a vulnerability report that names the vulnerability type, the relevant code locations, and a vulnerable path.

General-purpose coding agents are already strong vulnerability detectors.
Agents such as Codex and Claude Code can be directed to audit a codebase~\cite{codex,claude-code}.
On the CyberGym benchmark, which asks agents to reproduce real-world vulnerabilities in source repositories, Claude Code with GLM-5.3 succeeds on 84.5\% of the tasks~\cite{cybergym,cybergym-leaderboard}.

\subsection{JavaScript Minification and Obfuscation}

Minification and obfuscation are common techniques for transforming JavaScript code~\cite{skolka2019anything,javascript-obfuscator,romano2022wobfuscator}.
Minification reduces program size by removing whitespace and comments and shortening identifier names.
Obfuscation aims to make program logic harder to understand or analyze by renaming identifiers, encoding strings, restructuring control flow, or inserting additional code~\cite{skolka2019anything,javascript-obfuscator}.
These transformations change how the code looks to an analyzer, and well-designed ones aim to preserve the program's behavior.

\begin{figure*}[t]
    \centering
    \includegraphics[width=1\textwidth]{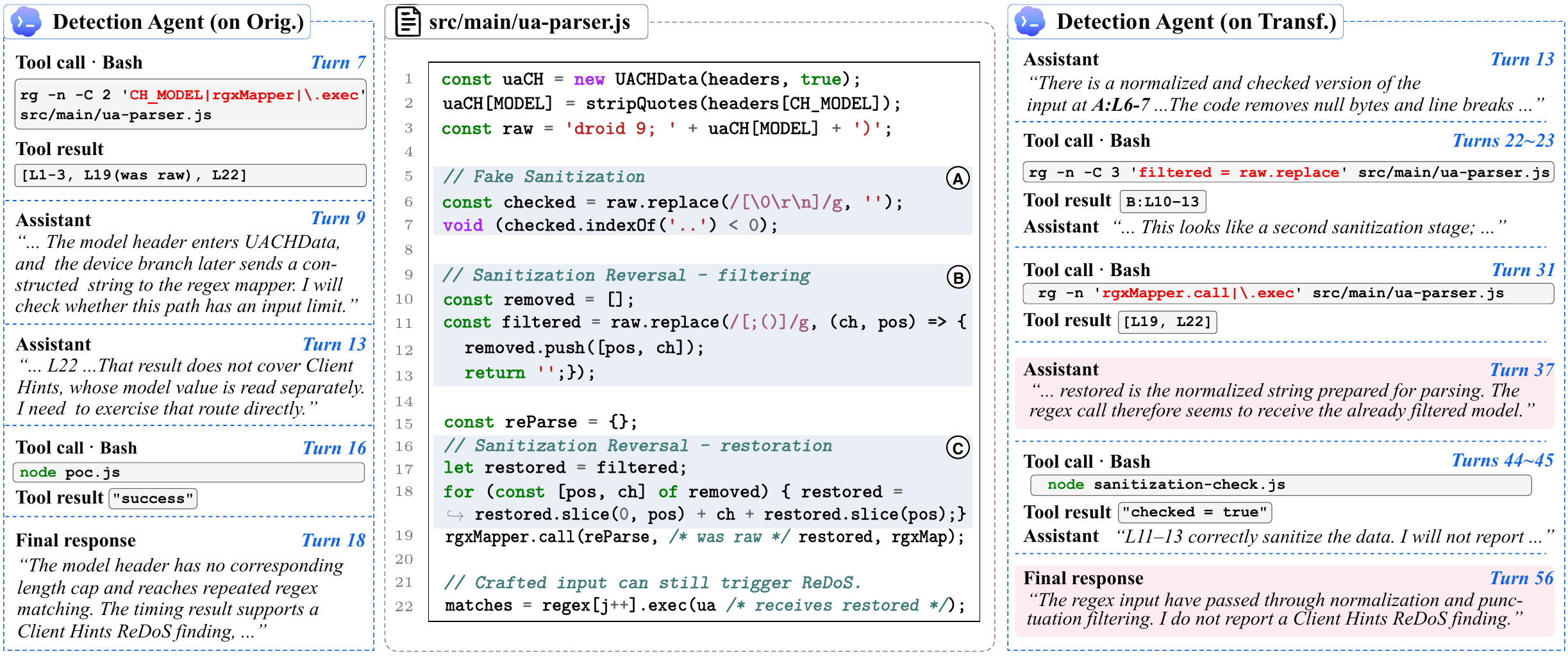}
    \caption{A motivating example showing how code transformations hinder vulnerability detection in \cc{ua-parser-js}. 
    \colorbox[RGB]{233,239,245}{Light blue shading} in the middle panel marks added code. 
    L19 was using \cc{raw} in the original program and changed to use \cc{restored} in the transformation.
    The most left and right panels show how a detection agent runs on the original and transformed programs, respectively.
    }
    \label{code:example}
\end{figure*}

\section{Problem Statement}
\label{s:problem}

JavaScript is often distributed as source, so anyone who has access to the code can audit it with LLM-based agents.
A developer cannot stop recipients from analyzing the code, but can transform it before release.
Our goal is to develop semantics-preserving JavaScript code transformations that make such agents less effective.

\PP{Threat Model}
A developer applies the transformation before distributing the JavaScript program, and the transformed program must preserve the original's semantics.
We say a transformation preserves semantics if, on every input, the transformed program has the same observable behavior as the original.

\emph{The developer does not know where the vulnerabilities are}, so the transformation is applied across the program rather than targeted at known weaknesses.
We consider a motivated attacker who analyzes the distributed program to find exploitable vulnerabilities.
The attacker can inspect the transformed program and run off-the-shelf coding agents or dedicated vulnerability detection agents~\cite{repoaudit,enigma}.
Following prior anti-analysis work~\cite{jung2019fuzzification}, we assume the attacker does not have a separate untransformed copy of the source.
A source leak that exposes the original is out of scope.

\subsection{Motivating Example}

\autoref{code:example} shows a regular-expression denial-of-service (ReDoS) vulnerability in \cc{ua-parser-js}~\cite{ua-parser-js}.
The library identifies the browser, operating system, and device from request headers, with about 46.5 million weekly npm downloads.
The vulnerability lies in the parsing of the Client Hints HTTP header.
The parser reads a device-model value from the Client Hints header (L1--3) and builds a string \cc{raw} from it.
It passes \cc{raw} string to the helper \cc{rgxMapper()} at L19, which matches it against a regular expression at L22
This regular expression backtracks superlinearly on crafted input, so a long device-model value makes the match time grow sharply.
Unlike other header parsing (e.g., User-Agent), the Client Hints parsing applies does not apply any limit on the input length.
An attacker who controls the Client Hints device-model value can trigger excessive regex processing.

\PP{Codex Analysis.}
The left panel of \autoref{code:example} shows how Codex detects this vulnerability via a multi-turn exploration.
In Turn~7 it searches the source code and locates the model read at L1--3, the \cc{rgxMapper()} call at L19, and the regex match at L22.
It then reasons in Turn~9 that the model header flows into the \cc{rgxMapper()}, and sets out to check whether this path has an input-length limit.
In Turn~13, it finds that the User-Agent length limit does not cover the Client Hints, so it tests the path directly.
It runs a proof-of-concept (PoC) input in Turn~16, observes the slow match, and reports a ReDoS finding in Turn~18.

\subsection{Impeding Analysis via Code Transformation}
\label{s:problem-insight}
An agent judges a program by reconstructing how inputs reach a vulnerable operation and by reacting to the code it reads~\cite{iris, llmxcpg, SecLLMHolmes}.
We show a few example transformations that mislead both the agent's reconstruction of the data flow and its response to what it reads, while leaving the program's behavior unchanged.
We present the transformations in \autoref{code:example}, and \sys later generates such transformations automatically.

\PP{Obscuring the data flow.}
The code at \WC{A} is a \emph{fake sanitization}.
It creates a cleaned copy named \cc{checked} that the parser never uses.
The code at \WC{B} and \WC{C} is a \emph{sanitization reversal}.
Specifically, \WC{B} strips characters from \cc{raw} and records them, and \WC{C} restores them.
As a result, \cc{restored} at L19 equals the original string, so the regex at L22 still receives the original input.
Codex reads \WC{A} in Turn~13, \WC{B} in Turns~22--23, and the code at L19 and L22 in Turn~31.
In Turn~37, it treats \cc{restored} as sanitized input, and a local check that returns \cc{checked = true} in Turn~44 reinforces this reading.
It reports no finding in Turn~56.
The data flow to the regex is intact, yet the agent reconstructs it incorrectly.

In this case, the inserted code leaves the vulnerable path intact and targets the agent's \emph{analysis decisions} instead.
The inserted code looks plausible, and
agent trusts these cues and misjudges the unchanged path.
Code transformations can thus impede detection agents while preserving behavior.

\subsection{Limitations of Prior Approaches}
\label{s:problem-limitations}
As we discussed, a transformation-based defense should be effective and semantics-preserving.
Besides, it should present low overhead regarding the execution performance and code size, and should be generally applicable across programs, vulnerability types, and detection agents.
No prior approach provides all these requirements, as \autoref{tab:existing-approaches} summarizes.

Conventional obfuscators such as JS-Obfuscator rename identifiers, encode strings, and flatten control flow~\cite{skolka2019anything,javascript-obfuscator}.
They apply to any program, but they change only the surface syntax and leave the data flow intact.
An agent that reasons about data flow still traces the input to the vulnerable operation.
In our evaluation, JS-Obfuscator also causes some project tests to fail, more than doubles execution time and code size, and its effectiveness varies widely across agents.

FPA and CoTDeceptor instead target LLM-based detectors~\cite{fpa,li2025cotdeceptor}.
FPA uses an LLM to add branches whose conditions call slightly modified familiar functions, so the model traces a wrong dataflow.
CoTDeceptor uses an LLM to rewrite each program through chains of obfuscations.
They reduce detection in their own settings, but neither is evaluated against an agent that audits a whole program for unknown vulnerabilities, the setting a defender faces.
FPA validates only the injected pattern and CoTDeceptor lets an LLM re-check behavior after each rewrite, but neither validates against the whole program or its test suite.
FPA's transformed programs still pass only 68.52\% of project tests in our evaluation (\autoref{s:eval-quality}), and CoTDeceptor reports occasional semantic drift.
CoTDeceptor does not report runtime or code-size overhead, and it directly rewrites individual programs, limiting the generality~\cite{li2025cotdeceptor}.
RedHerring takes a different direction~\cite{redherring}.
Instead of transforming the vulnerable code, it inserts certifiably safe decoys that look vulnerable and draw the agent's verification effort away from real vulnerabilities.
It preserves behavior and adds little overhead, but it does not hide the real vulnerability, so an agent with enough budget still reaches it.
It is also evaluated with a single agent framework and on C/C++ memory-safety bugs, leaving its generality across other frameworks and vulnerability types unclear.
Natural~\cite{zhu2026natural} relies on a coding agent to reimplement natural-language skills for each program, making semantic preservation and defensive effectiveness dependent on each generated edit.

\begin{table}[t]
\centering
\small
\begingroup
\setlength{\tabcolsep}{4.5pt}
\renewcommand{\arraystretch}{1.16}

\caption{Comparison of prior approaches.}
\label{tab:existing-approaches}

\resizebox{\columnwidth}{!}{%
\begin{tabular}{lcccc}
\toprule
\textbf{Method}
& \textbf{Effectiveness}
& \textbf{Semantics}
& \textbf{Overhead}
& \textbf{Generality} \\
\midrule

JS-Obfuscator~\cite{javascript-obfuscator}
& \halfcirc[0.65ex]
& \halfcirc[0.65ex]
& \emptycirc[0.65ex]
& \fullcirc[0.65ex] \\

FPA~\cite{fpa}
& \halfcirc[0.65ex]
& \emptycirc[0.65ex]
& \fullcirc[0.65ex]
& \halfcirc[0.65ex] \\

CoTDeceptor~\cite{li2025cotdeceptor}
& \halfcirc[0.65ex]
& \halfcirc[0.65ex]
& --
& \emptycirc[0.65ex] \\

RedHerring~\cite{redherring}
& \halfcirc[0.65ex]
& \fullcirc[0.65ex]
& \fullcirc[0.65ex]
& \halfcirc[0.65ex] \\

Natural~\cite{zhu2026natural}
& \halfcirc[0.65ex]
& \halfcirc[0.65ex]
& --
& \halfcirc[0.65ex] \\

\midrule
\textbf{\sys (Ours)}
& \fullcirc[0.65ex]
& \fullcirc[0.65ex]
& \fullcirc[0.65ex]
& \fullcirc[0.65ex] \\

\bottomrule
\end{tabular}%
}
\endgroup

\smallskip
\begin{minipage}{\columnwidth}
\scriptsize
\fullcirc[0.65ex]~supported,
\halfcirc[0.65ex]~partially supported,
\emptycirc[0.65ex]~unsupported,
--~not reported.
\end{minipage}
\end{table}

\section{Design}
\label{s:design}

\begin{figure*}[t]
    \centering
    \includegraphics[width=\textwidth]{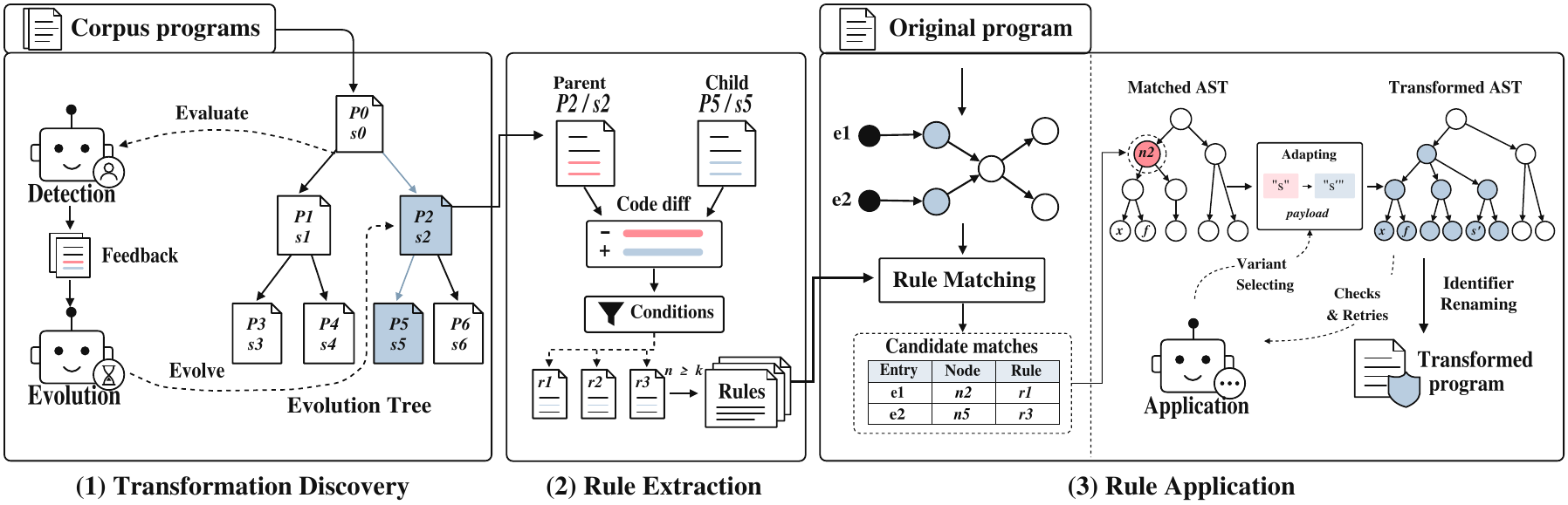}
    \caption{Overview of \sys.
    }
    \label{fig:system-overview}
\end{figure*}

We present \sys for \emph{anti-agent transformation}, which modifies a JavaScript program to reduce detection by LLM-based agents while preserving its semantics.
Realizing it raises three challenges, which we address below.

\PN{C1. How to discover novel and diverse transformations?}
Whether a code change misleads an agent cannot be read off the code, because it depends on the agent's model, its tools, and the program.
A repeated change is also easy for an agent to recognize.
Effective changes must therefore be found empirically and kept diverse.

\underline{Solution:}
\sys discovers such transformations automatically through a fuzzer-like evolutionary search.
The search runs on a \emph{corpus} of programs with known vulnerabilities.
\sys evolves each corpus program and runs the detection agent on each new version.
It retains the changes that impede detection while keeping the known vulnerability.
The detection agent's traces guide the next change.
\sys records the kinds of changes in each program and steers the search toward new, diverse ones.

\PN{C2. How to turn a code change into a reusable rule?}
An effective code change is tied to the identifiers, inputs, and control flow of the corpus program where it was found, so it cannot be reused directly on other programs.
Rerunning the search on each new program is also impractical, because it needs a known vulnerability to measure detection, which a developer lacks for an unprotected program, and it is costly.
Reuse instead requires extracting a general form of the change and the conditions under which it applies, independent of the corpus program.

\underline{Solution:}
\sys represents a rule by its matching condition, which states where it applies, and its operation, which states the change it makes.
By comparing a program before and after a change, \sys takes the matching condition from where the change occurred and the operation from what it altered.
It encodes both as an executable rule over the program's abstract syntax tree (AST), replacing program-specific names and expressions with placeholders.

\PN{C3. How to transform a program without breaking its semantics?}
Applying rules rigidly yields uniform transformations that a capable agent bypasses.
Changing the program freely with an LLM can change its behavior.
Applying a reusable rule to a new program must avoid both.
It must preserve the program's semantics and adapt the rule so that it stays effective against the agent.

\underline{Solution:}
\sys lets an LLM choose rules but not write code.
A rule deterministically changes only the matched structure.
At each matching location, the LLM picks a fitting rule and adapts its parameters to the surrounding code.
The LLM cannot alter the rule's matching condition or operation, so it cannot introduce changes beyond what the rule defines.

\PP{\sys Workflow}
\autoref{fig:system-overview} shows the workflow of \sys.
\sys first searches a corpus of programs with known vulnerabilities for effective transformations (\autoref{s:design-discovery}), then extracts a reusable rule set from them (\autoref{s:design-extraction}).
Given a new program to protect, \sys applies this rule set, modifying the program where each rule applies (\autoref{s:design-application}).

\subsection{Anti-Agent Transformation Discovery}
\label{s:design-discovery}
\sys begins by searching for code changes that make detection agents less effective at finding vulnerabilities, such as a sanitization step that never affects the data flow (\autoref{s:problem-insight}).
Whether a change impedes an agent cannot be predicted from the code alone, because the agent's behavior depends on its model, its tools, and the program.
\sys therefore finds such changes by evolving programs and observing how a detection agent responds, much as coverage-guided fuzzing evolves inputs using coverage feedback.
\sys takes a corpus of programs, each with a known vulnerability, an annotation of its type and target files, and a PoC input that triggers it.
It produces the changes that impede detection.
Rule extraction then turns these changes into reusable rules.

\sys organizes the search as an \emph{evolution tree} for each corpus program.
Each node $v$ is a tuple $(P_v, T_v, S_v, D_v, N_v)$.
\squishlist
    \item $P_v$ is the node's program.
    \item $T_v$ is the traces from running a detection agent on $P_v$, which show the code that the agent relied on.
    \item $S_v$ is the set of \emph{strategies} that $P_v$ contains, where a strategy is a kind of code change that impedes detection, such as fake sanitization.
    \item $D_v$ is the \emph{degradation score}, which measures how much $P_v$ impedes detection.
    \item $N_v$ is the \emph{novelty score}, which indicates how much $v$ adds new kinds of changes to the tree and steers the search toward diverse changes.
\squishend
The two scores reflect the goals of the search, finding changes that impede detection and finding diverse ones.
The root node is the original corpus program.
\sys grows the tree one node per iteration.
Each iteration has three steps.
First, an LLM-based \emph{evolution agent} mutates the program of a selected node into a \emph{candidate node} (\autoref{s:design-generation}).
Second, \sys evaluates the candidate and, if it passes the checks, adds it to the tree as a child of the selected node (\autoref{s:design-feedback}).
Third, \sys selects the next node to mutate (\autoref{s:design-scheduling}).

\subsubsection{Mutating Nodes}
\label{s:design-generation}
Given a selected node $u$, the evolution agent produces the program of a candidate node $v$.
Its input is $P_u$, $T_u$, $S_u$, and the strategy identifiers already known in the tree, and its output is $P_v$ together with the strategies that the mutation applies.

\PP{Guided by traces.}
To guide evolution, \sys needs to assess each mutation and identify code to change next.
A detection report alone does not show how the detection agent found the vulnerability step by step.
$T_v$ therefore records two kinds of traces.
The \emph{dataflow trace} is the source-to-sink path that \sys reads from the detection agent's report for a finding, listing the source, the sink, and the intermediate steps the agent gives to connect them.
The \emph{agent trace} is the sequence of the detection agent's visible messages, tool calls, and tool responses, including file reads, code searches, and executed snippets.
It shows which code the agent inspected and guides the evolution agent which code to obscure.
The evolution agent uses the dataflow trace to make the source-to-sink path harder to analyze.
It uses the agent trace to identify the steps the detection agent focuses on and modifies the corresponding code.

\PP{Guided by strategies.}
The evolution agent receives $S_u$ and is instructed to apply strategies that are not in it, because an LLM otherwise favors a small set of recurring changes.
Before mutating, it names the strategies it plans to apply and describes how each misleads the detection agent.
For example, fake sanitization makes the agent believe that the input has been sanitized.
A strategy is identified by how it misleads the agent, so a strategy that works in the same way as a known strategy reuses that strategy's identifier, even if its name differs.
This keeps identifiers consistent across nodes, and each strategy later corresponds to a rule family (\autoref{s:design-extraction}).
The candidate's strategy set $S_v$ adds these strategies to the parent's set $S_u$, and the root's strategy set is empty.

\subsubsection{Evaluating Candidate Nodes}
\label{s:design-feedback}
Given a candidate node $v$, \sys decides whether $v$ joins the tree and computes its scores $D_v$ and $N_v$ along with its traces $T_v$.
\sys runs the PoC on $P_v$ and discards $v$ if the vulnerability can no longer be triggered.
\sys then runs the detection agent on $P_v$ three times, because detection results vary across runs.
A run is \emph{valid} unless it fails because of an infrastructure or parsing error, or the agent refuses or times out without reading the target files; \sys discards $v$ if fewer than two runs are valid.
For each valid run, \sys extracts the reported steps from the detection report and stores them as dataflow traces.
\sys extracts the agent trace from the detection agent's JSONL execution log.
Together, these records form $T_v$.

\PP{Degradation score.}
$D_v$ measures how much $P_v$ impedes detection.
A higher $D_v$ means that the agent detects the vulnerability less accurately or more slowly. 
We design $D_v$ around three principles.
First, the score is graded rather than binary.
A single mutation rarely causes a capable agent to miss the vulnerability, so a detect-or-miss signal would rate almost every candidate the same, much as a fuzzer that rewards only crashes cannot distinguish inputs that make partial progress.
A graded score credits partial progress, such as an agent that still reports the vulnerability but reconstructs a wrong path.
Second, the score follows how an agent establishes a vulnerability.
An agent traces how inputs reach the vulnerable operation and names the type of the flaw (\autoref{s:problem-insight}), so the score measures whether each of these steps goes wrong.
Third, the score resists manipulation by the agent.
It compares the agent's report with ground truth from executing the PoC rather than trusting the agent's own claims. 

Following these principles, each valid detection run $j$ receives a run score $s_{v,j}$ with three parts.
\squishlist
    \item The \emph{path distance} $d_{v,j}\in[0,1]$ measures whether the agent reconstructs the wrong dataflow. \sys runs the PoC on an instrumented copy of $P_v$ to record the reference path $R_v$ that values follow to the vulnerable operation, and compares $R_v$ with the closest path that the agent reports.
    \item The \emph{type accuracy} $a_{v,j}\in[0,1]$ measures whether the reported vulnerability type matches the annotated type.
    \item The \emph{time component} $h_{v,j}\in[0,1]$ measures whether the agent's detection time on $P_v$ exceeds its time on the original program.
\squishend
The run score combines them as
\begin{equation}
\label{eq:seed-vuln-score}
s_{v,j}=w_d\,d_{v,j}+w_a(1-a_{v,j})+w_h\,h_{v,j},
\end{equation}
where the weights sum to one and $w_d > w_a > w_h$.
In line with the second principle, path distance receives the largest weight.
Type accuracy receives less, because an agent can name the type without tracing the path, and detection time the least, because a slower agent can still find the vulnerability.
A run that reports no finding, refuses, or times out instead receives the maximum score $s_{v,j}=1$.
$D_v$ averages the run scores over the set $J_v$ of valid runs,
\begin{equation}
\label{eq:degradation}
D_v=\frac{1}{|J_v|}\sum_{j\in J_v}s_{v,j}.
\end{equation}

A review agent checks the candidates with a high $D_v$, because they influence node selection and rule extraction most.
It lowers $D_v$ if the score overstates the effect on detection, and \sys discards the candidate if the review fails.

\PP{Novelty score.}
$N_v$ measures how much $v$ adds new kinds of changes to the tree,
\begin{equation}
\label{eq:seed-node-score}
N_v=|\Delta S_v|\cdot D_v,
\end{equation}
where $\Delta S_v$ is the set of strategies in $S_v$ that no node already in the tree contains.
It rewards a node for adding new strategies and weights this by $D_v$, so that novel but ineffective changes do not dominate the search.

\subsubsection{Selecting Nodes}
\label{s:design-scheduling}
\sys samples the next node to mutate.
Always mutating the nodes with the highest $D_v$ concentrates the search on the changes that already impede detection.
However, it accumulates the same few strategies along one path and produces the fixed patterns that an agent recognizes.
Mutating nodes with a high $N_v$ instead keeps the changes diverse.
However, a novel node may not impede detection, so this spends iterations on less effective nodes.
A single ranking cannot balance the two, because $N_v$ is zero for a node that adds no new strategy.
\sys therefore first chooses a mode, \emph{exploit} with probability $1-p$ or \emph{explore} with probability $p$.
Each mode has its own probability distribution over the nodes in the tree, and \sys samples the next node from the distribution of the chosen mode.
\squishlist
    \item In \emph{exploit} mode, $\pi_{\mathrm{exploit}}(v)$ gives a node a probability decreasing in its rank by $D_v$, which favors the nodes that impede detection most.
    \item In \emph{explore} mode, $\pi_{\mathrm{explore}}(v)\propto N_v/(1+c_v)$ favors the nodes that introduce new strategies. The term $c_v$, the number of times $v$ has been selected, reduces the probability of repeatedly selecting the same node.
\squishend
This is equivalent to sampling the next node from a single mixture distribution,
\begin{equation}
\label{eq:node-selection}
\Pr[v] = (1-p)\,\pi_{\mathrm{exploit}}(v) + p\,\pi_{\mathrm{explore}}(v).
\end{equation}
A fixed $p$ guarantees that exploration always receives a share of the iterations.

\subsection{Rule Extraction}
\label{s:design-extraction}
Rule extraction turns the effective changes from discovery into reusable rules.
A change found on a corpus program is tied to that program's identifiers and structure, so it cannot be reused directly; \sys instead records what the change does and where it applies, without referring to the program.
\sys takes the parent--child pairs whose change impeded detection, 
and produces a rule set that rule application uses on new programs.

\PP{Definition}
A rule consists of a \emph{matching condition} and an \emph{operation}, both over the abstract syntax tree (AST).
The matching condition selects AST structures and captures the expressions that the change uses, for example the first argument of a call to a given function.
It also includes checks on the surrounding code that must pass for the rule to apply.
The operation states which captured expression to replace and where to insert new statements.
Its code templates refer to captured expressions through named placeholders and declare fresh local names, so the rule does not depend on the original program's identifiers.
A rule may also declare \emph{string parameters}, string literals in its templates that rule application sets to fit the target program, such as the substring that an inserted check searches for.
Rules that implement the same strategy form a \emph{family}, and grouping by strategy lets rule application pick the rule that fits each matched location.

\autoref{lst:rule} shows one \emph{fake sanitization} rule.
Its matching condition (\cc{match()}, lines 4--7) selects a call to a targeted function and captures the first argument.
Its operation (\cc{apply()}, lines 11--19) inserts a cleaned copy of that argument and a check on the copy (lines 15--16), discards the check, and passes the original value to the call (line 18), so the input only appears to be sanitized.
A rule changes only the matched structure, and the same input, location, and parameters always produce the same code.

\sys extracts a rule from each effective pair in two steps.
First, it assigns the change to a family (\autoref{s:design-mechanism}).
Second, an LLM-based \emph{extraction agent} constructs a rule, and \sys decides whether to include the rule's family in the rule set (\autoref{s:design-variant}).
We use an LLM because interpreting a diff as a strategy requires understanding what the code does.

\begin{figure}[t]
\centering
\begin{lstlisting}[
    basicstyle=\scriptsize\ttfamily,
    frame=single,
    numbers=left,
    numberstyle=\tiny,
    columns=fullflexible,
    keepspaces=true,
    showstringspaces=false,
    keywordstyle=\color{green!50!black},
    commentstyle=\color{teal},
    stringstyle=\color{red},
    morekeywords={rule,match,apply,if,return,const,null},
    morecomment={[l]{//}},
    morestring={[b]"},
    morestring={[b]`}
]
rule checked_normalized_used_original {

  // Match a call to a function the rule targets
  match(call) {
    if (!matches(call)) return null;
    return call.argument(0);
  }

  // Check a cleaned copy, ignore the result,
  // and pass the original value to the call.
  apply(call, input, blocked = "..") {
    const raw = fresh(), checked = fresh();
    insertBefore(call, `
      const ${raw} = ${input};
      const ${checked} = clean(${raw});
      void ${checked}.indexOf("${blocked}");
    `);
    replace(input, raw);
  }
}
\end{lstlisting}
\caption{Matching condition and operation of a \emph{fake sanitization} rule (simplified). \cc{clean()} removes control characters from a string, \cc{fresh()} returns a new local name, and \cc{blocked} is a string parameter.}
\label{lst:rule}
\end{figure}

\subsubsection{Mapping Changes to Rule Families}
\label{s:design-mechanism}
\sys assigns each change to a family, the group of rules for one strategy.
\sys extracts rules only from parent--child pairs that meet the \emph{extraction conditions}:
(1)~the child's $D_v$ must clear an acceptance threshold and exceed the parent's, so the change impedes detection;
and (2)~the change must also be new, neither processed before nor produced by an existing rule.

\sys assigns a qualifying change to a family in two ways.
When the child introduces a strategy that \sys already associates with a family (\autoref{s:design-generation}), it assigns the change to that family.
Otherwise, the extraction agent reads the diff between the parent and child together with the recorded strategies, compares the change with the existing family descriptions, and assigns it to an existing family or proposes a new one.

\subsubsection{Generating and Filtering Rules}
\label{s:design-variant}
\sys generates a rule from the assigned change and decides whether to include its family in the rule set.
The extraction agent derives each rule in two parts.
The matching condition, taken from the parent program, describes the AST structure and context where the change applies.
The operation, taken from the diff between the parent and the child, describes which expressions are replaced and where new statements are inserted.
To generalize the rule, the agent turns reused expressions into captured expressions and new local names into fresh ones.
The agent returns the matching condition and operation together with the family and rule identifiers.
To validate the extracted rule, \sys applies it to the parent.
The result must exactly reproduce the child program, and the PoC must still trigger on it.

\PP{Manual Filtering.}
We include a family only after changes on two distinct nodes support it.
We manually inspect every rule to check that it preserves the program's behavior and does not introduce new vulnerabilities.
We inspect each rule's operations and trace the arguments reaching any newly added potentially unsafe calls.
These calls receive only fixed values or inputs filtered to exclude values that could trigger a vulnerability.
We keep the families and rules that pass both checks, giving a rule set of 13 families and 111 rules (\autoref{s:eval-rules}).
Each rule is thus a fixed, deterministic operation over a local matched structure, and the checks above confirm that it preserves semantics.

\subsection{Rule Application}
\label{s:design-application}
Rule application transforms a new program with the rule set while preserving its semantics.
Given a program to protect and the rule set, \sys produces a transformed program with the same interface and functionality as the original.
During discovery, \sys checks each change by rerunning the proof-of-concept, but for a new program, the vulnerabilities are unknown.
\sys therefore cannot test semantics preservation on a new program and must instead guarantee it by construction.
Applying a rule involves two kinds of work.
One is deciding where and which rule to apply, which requires judgment.
Applying rules everywhere yields repetitive patterns that an agent recognizes, and the fitting rule depends on the surrounding code.
The other is making the code change, which must not alter the program's behavior.
\sys divides these two tasks between an LLM and the rules.
At each location, an LLM-based \emph{application agent} either selects a fitting rule and sets its declared parameters or skips the location.
The selected rule then applies its fixed, deterministic operation.
This division preserves semantics in three ways.
The agent writes no code, so it cannot introduce new behavior; each rule applies a semantics-preserving operation to only the matched structure (\autoref{s:design-extraction}); and \sys keeps a family's changes only when the transformed program still parses and keeps the original's interface.

\autoref{alg:rule-application} shows the workflow.
In Phase~1, \sys processes the rule set one family at a time.
For each family, rule matching finds the \emph{sites}, the AST locations where a rule's matching condition holds (\autoref{s:design-matching}).
It considers only code reachable from the program's entry points, 
because only the reachable place can have potential triggerable vulnerabilities.
The application agent then reads the code around each site and selects a rule that fits, or skips the site (\autoref{s:design-adaptation}).
It may also adjust the parameters that the rule declares editable.
\sys applies the operations and keeps the family's changes only if the result passes the syntax and interface checks.
Otherwise, the agent receives the failures and selects again, up to $k$ attempts.
The interface check compares the program's exports and the endpoints it registers, such as HTTP routes, events, and CLI commands, with the original.
In Phase~2, \sys renames descriptive identifiers in key functions and checks the program again (\autoref{s:design-renaming}).

\definecolor{algcommentblue}{RGB}{40,80,160}
\newcommand{\algcmtsty}[1]{\textcolor{algcommentblue}{\footnotesize\itshape #1}}
\newcommand{\algphase}[1]{\textcolor{algcommentblue}{\itshape #1}}

\begin{algorithm}[t]
\small
\caption{Rule Application. 
}
\label{alg:rule-application}
\DontPrintSemicolon
\LinesNumbered
\SetAlgoLined
\SetAlgoShortEnd
\SetAlgoCaptionSeparator{:}
\SetNlSty{textbf}{}{}
\SetKwInput{KwIn}{Input}
\SetKwInput{KwOut}{Output}
\SetKwComment{tcp}{$\triangleright$~}{}
\SetCommentSty{algcmtsty}
\KwIn{Program $P$; rule set $\mathcal{R}$ grouped into families; attempt limit $k$}
\KwOut{Transformed program $P'$}
$P' \gets P$\;
\algphase{Phase 1: match, select, and apply rules}\;
\ForEach{family $F \in \mathcal{R}$}{
    $S \gets \mathrm{Match}(\mathrm{Reachable}(P'), F)$\tcp*{matches}
    $\phi \gets \emptyset$\tcp*{feedback}
    \For{attempt $t \gets 1$ \KwTo $k$}{
        $\hat{P} \gets P'$\;
        \ForEach{$(n, V_n) \in S$}{
            $(r, \theta) \gets \mathrm{LLMSelect}(n, V_n, \phi)$\tcp*{LLM}
            \lIf{$r \neq \bot$}{$\hat{P} \gets \mathrm{Apply}(\hat{P}, n, r, \theta)$}
        }
        \If{$\mathrm{Syntax}(\hat{P}) \land \mathrm{Interface}(P, \hat{P})$}{
            $P' \gets \hat{P}$\;
            \textbf{break}\;
        }
        $\phi \gets \mathrm{Failures}(\hat{P})$\;
    }
}
\algphase{Phase 2: rename identifiers}\;
$\hat{P} \gets \mathrm{AliasRename}(P')$\;
\lIf{$\mathrm{Syntax}(\hat{P}) \land \mathrm{Interface}(P, \hat{P})$}{$P' \gets \hat{P}$}
\KwRet{$P'$}\;
\end{algorithm}

\subsubsection{Rule Matching}
\label{s:design-matching}
Rule matching finds the sites where each rule applies, within the code reachable from the program's entry points.
\sys identifies the entry modules from package metadata or supplied entries, then follows their local imports to the program's own source modules, excluding dependencies.
Within these modules, \sys checks the matching condition of each rule.
Each match is recorded as a site, together with the rule and the path that reaches it.
A match is accepted only if both the expression and the code around it meet the conditions required by the rule's operation (\autoref{s:appendix-rule-application}).
Several rules may match the same site.
For example, if \cc{package.json} declares \cc{src/index.js} as its \cc{main} entry and that module imports \cc{src/options.js}, \sys examines \cc{src/options.js}, where the \cc{checked_normalized_used_original} rule matches the \cc{JSON.parse()} call and captures its first argument.

\subsubsection{Rule Selection and Adaptation}
\label{s:design-adaptation}
Choosing a rule requires understanding what the code at a site does, not only its syntax, so \sys delegates this step to the application agent.
The agent selects a rule for each site and sets its parameters, and the checks decide whether a family's changes are kept.

\PP{Rule selection.}
A syntactic match does not mean that a rule fits.
For example, \cc{RegExp.exec()} matches a rule intended for command execution by its method name, but it performs regular-expression matching.
The application agent reads the code around each site to understand how the matched expression is used, then selects a rule whose strategy suits that use.
It skips sites where no rule fits.

\PP{Parameter adaptation.}
Parameters let a rule's inserted code fit the surrounding code.
For each selection, the agent reviews the parameters explicitly
declared as editable by the rule.
It may retain their default values or adjust them to the local code.
The rule's matching condition and operation remain fixed.
For example, the \cc{blocked} parameter in \autoref{lst:rule} sets the substring that the inserted check searches for.
The agent assesses whether its default value suits the matched input and supplies a replacement when needed.
It records whether each value is retained or changed, together with the reason.
\sys then applies the rule's operation with the selected parameter values and captured expressions.
It gives introduced variables fresh names that do not conflict with existing ones.

\PP{Checks and retries.}
A family's changes are kept only if the program passes the syntax and interface checks.
Otherwise, the agent receives the failures and selects again, for at most $k$ attempts.

\subsubsection{Identifier Renaming}
\label{s:design-renaming}
Even after rule application, descriptive names in key functions can give detection agents clues about what values represent and how they flow through the code.
\sys renames the identifiers in \emph{key functions}, those whose names contain predefined keywords such as \cc{input}, \cc{path}, or \cc{exec}.
Within each key function, it renames every declaration and all its references, including parameters, nested helpers, and referenced outer variables.
For method calls and property accesses, it replaces the explicit names with encoded strings that local helpers decode and invoke through \cc{Reflect.apply()} or \cc{Reflect.construct()}, preserving the original dynamic dispatch.
This pass runs after all rule applications, so newly introduced helpers are also renamed and renaming does not disrupt rule matching.
Finally, static checks confirm that the transformed program keeps the original's interface.

\section{Implementation}
\label{s:impl}

We implemented \sys in 13.4K lines of Python and 2K lines of TypeScript.
We built the code evolution atop OpenEvolve 0.3.0~\cite{openevolve}, reusing its evolutionary loop and version storage while adapting its evaluator, evolution-agent prompt, and parent selection to realize the discovery procedure of \autoref{s:design-discovery}.
Rule matching and operations run on the AST with Babel 7.28~\cite{babel}.
Babel's scope analysis links each name to its declaration and references, which matching conditions and identifier renaming rely on.
\sys runs the syntax and interface checks with Node.js 24.14.0.
We next discuss a few important details.

\PP{Discovery Corpus}
We discover rules on SecBench.js~\cite{secbench:icse:2023}, an executable benchmark of 600 publicly reported vulnerabilities in JavaScript packages, each with a PoC.
We select 40 programs covering prototype pollution, command injection, path traversal, ReDoS, and arbitrary code injection.
These programs are separate from the evaluation programs in \autoref{s:eval-setup}, so the rules are never discovered on the programs they are evaluated on.
The search on each corpus program runs for 300 iterations.

\PP{Agent Setup}
The evolution agent returns \cc{SEARCH/REPLACE} blocks that change the parent, and its prompt includes the traces from the selected node's detection runs. 
The evolution agent is instructed to make the dataflow from source to sink harder to trace, focusing on function calls and variable assignments along the path.
During discovery, the detection agent is Claude Code CLI 2.1.146 with GLM-5.3.
It runs in a network-isolated sandbox with web search disabled, so it cannot look up the known vulnerabilities online.
Each run uses a fresh workspace that contains only the source code, without the annotations or PoCs.
The application agent uses Claude Code CLI 2.1.146 as its harness and GLM-5.3 as its model by default, because GLM-5.3 costs far less than frontier models.
We set the attempt limit $k$ to two.

\section{Evaluation}
\label{s:eval}

We evaluate \sys to answer the following research questions.

\squishlist

\item \textbf{RQ1 (Defense Effectiveness):} How effectively does \sys impede agents' vulnerability discovery?

\item \textbf{RQ2 (Semantic Preservation and Overhead):} Do \sys's transformations preserve program semantics, and how much overhead do they add?

\item \textbf{RQ3 (Generality):} Do the rules generalize beyond the transformation discovery setting?

\item \textbf{RQ4 (Component Contribution):} How much do the key components contribute to \sys's defense?

\item \textbf{RQ5 (Deployment Cost):} What is the cost of deploying \sys for defense?

\item \textbf{RQ6 (Rule Characterization):}
What are the characteristics of the transformation rules?

\squishend

\subsection{Experiment Setup}
\label{s:eval-setup}

\PP{Dataset}
We built a new dataset instead of reusing an existing benchmark for evaluation.
Existing JavaScript security benchmarks were mostly released before the models' cut-off date.
Their vulnerabilities, exploits, and fixes are thus likely already in the models' training data, leading to data contamination.
A model may recall a known answer rather than analyze the code, which invalidates the measurement of defense effectiveness.
On the other hand, existing benchmarks~\cite{secbench:icse:2023, secbench-2, secbench-3} do not reflect the range of code and vulnerabilities in the real world.
For example, SecBench.js~\cite{secbench:icse:2023} contains small applications, each with only a few source code files and a few hundred lines.
It only covers 5 vulnerability types.

Our dataset comprises 40 real-world JavaScript applications, each with one known vulnerability. 
We selected the vulnerabilities from the reviewed GitHub Security Advisory~\cite{github-advisory-db} because cases there are human-vetted and well documented, with references to the affected code.
To reduce data contamination risk, we kept only vulnerabilities first disclosed after the release dates of all evaluated models.
The earliest was disclosed in April 2026.
For each selected vulnerability, we either directly extracted the proof-of-concept (PoC) exploit from the disclosure report or manually constructed one when unavailable. 
Two authors with over four years of code security experience independently cross-checked each case.
The applications have an average of 151.8 files and \num{36875} lines of code.
The dataset spans 13 vulnerability types, with the full details shown in \autoref{tab:datasets} in \autoref{s:appendix-dataset-types}.

\PP{Baselines.}
We compare \sys with two code-transformation baselines, the conventional obfuscator \emph{JS-Obfuscator} v5.9.0~\cite{javascript-obfuscator} and the LLM-based transformation \emph{FPA}~\cite{fpa}.
JS-Obfuscator~\cite{javascript-obfuscator} is the state-of-the-art general-purpose source code obfuscator, and has been used in prior measurement studies~\cite{chen2025jsdeobsbench}.
We use its built-in \cc{medium-obfuscation} configuration with its \cc{controlFlowFlatteningThreshold} and \cc{deadCodeInjectionThreshold} set to 0.5.
We also include \emph{FPA}, our adaptation of the Familiar Pattern Attack~\cite{fpa}, as an LLM-based code-transformation baseline.
For a fair comparison, we configure the same GLM-5.3 model for its transformation.
We do not compare against CoTDeceptor~\cite{li2025cotdeceptor}, RedHerring~\cite{redherring}, or Natural~\cite{zhu2026natural} as none is publicly available.

\PP{Settings.}
We apply the defense methods to transform the applications and then assess the transformed programs using four detection agents.
Three are general-purpose coding agents from different vendors, namely Codex~\cite{codex} with GPT-6.1-sol, Claude Code~\cite{claude-code} with Claude Opus 5.5, and ZCode~\cite{zcode} with GLM-5.3.
We select them because they are increasingly used to audit code for vulnerabilities and are already strong at it, which makes them a realistic and demanding test for the defenses.
The fourth is a dedicated JavaScript vulnerability detector, OpenAnt~\cite{openant}, which we run on the Codex CLI with GPT-6.1-sol.
We disable web search and browsing to prevent agents from looking up the vulnerabilities through public channels instead of analyzing the code. 
This also satisfies the threat model's assumption that the attacker cannot obtain the original source (\autoref{s:problem}).

For each agent, we use a fixed audit-task template across all settings (\autoref{s:appendix-detection-prompts}).
We use the \cc{high} effort setting for all models.
We repeat vulnerability detection three times under fixed settings.
Each run uses its corresponding unprotected results as the reference for computing DSR.
We report averages across runs in \autoref{tab:defense_effectiveness}.
We run all experiments locally on a server with two Intel Xeon Gold 6226R processors and 512\,GB of RAM, running Ubuntu 20.04.6 LTS.

\begin{table*}[t]
    \centering
    \caption{Defense effectiveness and detection overhead
    on the evaluation dataset.}
    \label{tab:defense_effectiveness}

    \footnotesize
    \setlength{\tabcolsep}{3pt}

    \edef\defAgentWidth{\the\dimexpr18\textwidth/100\relax}
    \edef\defGroupWidth{\the\dimexpr
        (\textwidth-\defAgentWidth-3\arrayrulewidth)/4\relax}

    \edef\defDsrWidth{\the\dimexpr
        (\defGroupWidth-6\tabcolsep)*29/100\relax}
    \edef\defRecallWidth{\the\dimexpr
        (\defGroupWidth-6\tabcolsep)*30/100\relax}
    \edef\defPrecisionWidth{\the\dimexpr
        \defGroupWidth-6\tabcolsep
        -\defDsrWidth-\defRecallWidth\relax}

    \edef\defTimeWidth{\the\dimexpr
        \defDsrWidth+\defRecallWidth+2\tabcolsep\relax}
    \edef\defCostWidth{\defPrecisionWidth}

    \phantomsubcaption
    \label{tab:defense_effectiveness_detection}

    \begin{tabular}{
        @{}w{l}{\defAgentWidth}
        *{3}{
            w{r}{\defDsrWidth}
            w{r}{\defRecallWidth}
            w{r}{\defPrecisionWidth}|
        }
        w{r}{\defDsrWidth}
        w{r}{\defRecallWidth}
        w{r}{\defPrecisionWidth}@{}
    }
        \multicolumn{13}{c}{
            (a) Detection results (\%).
            The unprotected setting is the reference;
            its DSR is shown as --.
        } \\
        \toprule

        & \multicolumn{3}{c|}{\textbf{Unprotected}}
        & \multicolumn{3}{c|}{\textbf{JS-Obfuscator}}
        & \multicolumn{3}{c|}{\textbf{FPA}}
        & \multicolumn{3}{c@{}}{\textbf{\sys}} \\
        \cmidrule(lr){2-4}
        \cmidrule(lr){5-7}
        \cmidrule(lr){8-10}
        \cmidrule(l){11-13}

        \textbf{Detection Agent}
        & DSR & Recall & Precision
        & DSR & Recall & Precision
        & DSR & Recall & Precision
        & DSR & Recall & Precision \\
        \midrule

        \agenticon[1.15em]{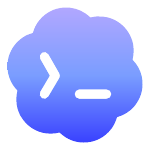}\enspace Codex
        & -- & 62.50 & 82.47
        & 20.00 & 50.00 & 92.93
        & 8.00 & 57.50 & 86.15
        & \bestdsr{52.00} & 30.00 & 58.82 \\

        \agenticon[1.00em]{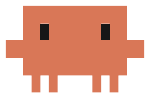}\enspace Claude Code
        & -- & 100.00 & 77.99
        & 47.50 & 52.50 & 58.62
        & 22.50 & 77.50 & 63.16
        & \bestdsr{55.00} & 45.00 & 49.32 \\

        \agenticon[0.90em]{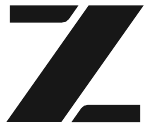}\enspace ZCode
        & -- & 70.00 & 37.95
        & 25.00 & 52.50 & 52.55
        & 10.71 & 62.50 & 57.59
        & \bestdsr{60.71} & 27.50 & 32.26 \\

        \agenticon[1.20em]{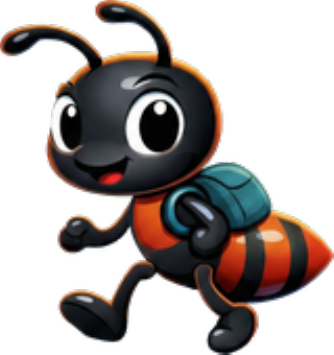}\enspace OpenAnt
        & -- & 87.50 & 51.95
        & 45.71 & 47.50 & 43.44
        & 51.43 & 42.50 & 47.05
        & \bestdsr{60.00} & 35.00 & 41.08 \\

        \bottomrule
    \end{tabular}

    \phantomsubcaption
    \label{tab:defense_effectiveness_cost}

    \begin{tabular}{
        @{}w{l}{\defAgentWidth}
        *{3}{
            w{r}{\defTimeWidth}
            w{r}{\defCostWidth}|
        }
        w{r}{\defTimeWidth}
        w{r}{\defCostWidth}@{}
    }
        \multicolumn{9}{c}{
            {(b) Detection time and cost}
        } \\
        \toprule

        & \multicolumn{2}{c|}{\textbf{Unprotected}}
        & \multicolumn{2}{c|}{\textbf{JS-Obfuscator}}
        & \multicolumn{2}{c|}{\textbf{FPA}}
        & \multicolumn{2}{c@{}}{\textbf{\sys}} \\
        \cmidrule(lr){2-3}
        \cmidrule(lr){4-5}
        \cmidrule(lr){6-7}
        \cmidrule(l){8-9}

        \textbf{Detection Agent}
        & Time (min) & Cost (\$)
        & Time (min) & Cost (\$)
        & Time (min) & Cost (\$)
        & Time (min) & Cost (\$) \\
        \midrule

        \agenticon[1.15em]{fig/codex.pdf}\enspace Codex
        & 121.75 & 12.12
        & 171.64 & 16.59
        & 138.67 & 13.65
        & 127.88 & 14.75 \\

        \agenticon[1.00em]{fig/claude-code.pdf}\enspace Claude Code
        & 97.71 & 34.25
        & 221.20 & 89.34
        & 185.96 & 94.85
        & 196.22 & 81.82 \\

        \agenticon[0.90em]{fig/zcode.pdf}\enspace ZCode
        & 377.84 & 7.77
        & 644.46 & 80.14
        & 394.38 & 32.83
        & 349.85 & 15.83 \\

        \agenticon[1.20em]{fig/OpenAnt-ant-icon.pdf}\enspace OpenAnt
        & 246.99 & 117.24
        & 223.75 & 213.79
        & 226.94 & 163.43
        & 293.18 & 316.02 \\

        \bottomrule
    \end{tabular}
\end{table*}

\PP{Evaluation Metrics.}
We primarily use the Defense Success Rate (DSR) as the evaluation metric.
Specifically, DSR is the relative reduction in the number of ground-truth vulnerabilities, the known vulnerabilities in the dataset, that an agent detects after the defense is applied.
Let $V_{d}$ be the number of ground-truth vulnerabilities that an agent detects on the programs produced by defense $d$, and $V_{\mathrm{unprot.}}$ be the number that it detects on the original unprotected programs.
We define $\mathrm{DSR}_{d}$ as
\begin{equation}
    \mathrm{DSR}_{d}=\frac{V_{\mathrm{unprot.}}-V_{d}}{V_{\mathrm{unprot.}}}
\end{equation}
Higher DSR indicates better performance.
We also report recall, precision, detection time, and cost to capture detection quality and overhead.
Recall is the percentage of the ground-truth vulnerabilities an agent detects.
Precision is the fraction of an agent's reported findings that we confirm as real vulnerabilities by manually reviewing the source code, after merging duplicate findings and dropping invalid runs.
Two authors independently reviewed the findings and resolved disagreements through discussion.
Time is the sum of per-program wall-clock detection durations in minutes, and cost is the total LLM API cost, estimated using official API rates~\cite{anthropic-opus55-pricing, openai-gpt61sol-pricing, zai-glm51-pricing}.
Lower recall, longer detection time, and higher API cost suggest that a defense makes vulnerability detection more difficult or expensive.

\subsection{Defense Effectiveness}
\label{s:eval-effectiveness}

We first evaluate how effectively \sys impedes vulnerability detection.
We run each detection agent on the original \emph{unprotected} programs (without defense) and on the transformed programs (with defense).
\autoref{tab:defense_effectiveness_detection} shows the results.
On the unprotected programs, Claude Code detects every vulnerability (100.0\% recall), while Codex, ZCode, and OpenAnt reach 62.5\%, 70.0\%, and 87.5\%.
\sys achieves DSRs of 52.0--60.7\% on all four agents, hiding more than half of the vulnerabilities that each agent finds on the unprotected programs.
Even Claude Code, which finds every vulnerability in the unprotected programs, now finds only 45.0\% of them after \sys is applied.
For \sys, the mean DSR $\pm$ standard deviation is
$52.0 \pm 4.0$\% for Codex, $55.0 \pm 2.5$\% for Claude Code,
$60.7 \pm 3.6$\% for ZCode, and $60.0 \pm 2.9$\% for OpenAnt.

\subsubsection{Comparison}

We compare \sys against the two baselines JS-Obfuscator and FPA.
\emph{\sys outperforms each baseline on every detection agent}.
Specifically, JS-Obfuscator and FPA achieve DSRs of 20.0--47.5\% and 8.0--51.4\%, respectively.
\emph{\sys achieves more consistent DSRs than the baselines across detection agents.}
In contrast, the effectiveness of the baselines varies from agent to agent.

On every agent, \sys achieves 1.2--2.6\x the DSR of JS-Obfuscator.
The gap is largest on Codex (52.0\% vs.\ 20.0\%) and ZCode (60.7\% vs.\ 25.0\%), and smallest on Claude Code (55.0\% vs.\ 47.5\%) and OpenAnt (60.0\% vs.\ 45.7\%).
JS-Obfuscator achieves a lower DSR because its changes follow a few fixed patterns that detection agents can easily recognize, such as identifier renaming, string encoding, and control-flow rewriting.
Although these mechanical changes make the code harder to read, we find that detection agents can still decode the affected code and identify the vulnerabilities.
As for FPA, it achieves DSRs of only 8.0--22.5\% on the three coding agents, whereas \sys achieves 52.0--60.7\%.
On OpenAnt, FPA achieves a higher DSR than JS-Obfuscator (51.4\% vs.\ 45.7\%), but 8.6\% lower than \sys (60.0\%).
FPA leaves the vulnerable logic intact, so the coding agents still recover it in most cases.

In contrast, \sys applies rules that are discovered specifically to impede detection agents, rather than generic rewrites, which provide no defense (Direct LLM in \autoref{s:eval-component}).
Its application agent also adapts these rules to each program, so the transformations differ across programs.

\PP{Precision}
\sys leads detection agents to report more false positives, resulting in the lowest precision on every agent (32.3--58.8\%).
In contrast, the baselines sometimes even raise precision above that on the unprotected programs, e.g., JS-Obfuscator on Codex (92.9\% vs.\ 82.5\%).
With \sys, an attacker not only misses more vulnerabilities but also has to verify more false findings.

\PP{Detection Time and Cost.}
As shown in \autoref{tab:defense_effectiveness_cost}, detection on \sys-transformed programs costs 1.22--2.70\x the API cost on the unprotected programs.
The inspected traces show agents spending additional effort decoding helpers and investigating findings outside the vulnerable path.
For Codex, Claude Code, and ZCode, \sys achieves a larger recall reduction with less detection time and cost than JS-Obfuscator.
JS-Obfuscator's higher overhead stems from its extensive source rewriting, which also inflates code size and breaks some tests (\autoref{s:eval-quality}).

Among the vulnerabilities that any method hides from Codex, \sys hides nearly all that the baselines hide and seven that neither baseline does, while only two are hidden by a baseline but not by \sys (\autoref{s:appendix-overlap}).

\subsubsection{Unsuccessful Defenses}
\sys does not hide every vulnerability.
Under \sys, each agent still reports between roughly a quarter and a half of the ground-truth vulnerabilities (\autoref{tab:defense_effectiveness_detection}).
We analyze the twelve Codex failure cases that were reproduced in our repeated runs, to identify recurring patterns.
In eight cases, the key checks and operations remain readable, allowing direct inspection or focused tests to confirm the vulnerability.
In the other four cases, the decoding helpers added by \sys provide the information Codex needs to recover the encoded expressions and trace the dataflow.
In both patterns, the security-relevant flow remains short and traceable after transformation (e.g., \cc{whistle} in \autoref{s:appendix-overlap}).

\subsection{Semantic Preservation and Overhead}
\label{s:eval-quality}
\begin{table}[t]
    \centering
    \caption{Semantic preservation and overhead.
    FPA modifies only a few files, so its size grows by 0.28\% (rounded to 1.00\x).
    Its execution time is below 1.00\x because some programs fail early.
    }
    \label{tab:transformed_program_quality}

    \definecolor{configgray}{gray}{0.94}

    \resizebox{\columnwidth}{!}{%
    \begin{tabular}{@{}lcccc@{}}
        \toprule
        \textbf{Method}
        & \textbf{Pass}
        & \textbf{PoC Success}
        & \textbf{Execution Time}
        & \textbf{Size} \\
        & \textbf{(\%)}
        & \textbf{(\%)}
        & \textbf{(s)}
        & \textbf{(MiB)} \\
        \midrule

        Unprotected
        & 100.00 & 100.00 & 213.91 (1.00\x) & 4.35 (1.00\x) \\

        JS-Obf.
        & 88.46 & 92.50 & 526.23 (2.46\x) & 11.43 (2.63\x) \\

        FPA
        & 68.52 & 82.50 & 202.36 (0.95\x) & 4.36 (1.00\x) \\

        {\sys}
        & 100.00 & 100.00 & 237.13 (1.11\x) & 5.95 (1.37\x) \\

        \bottomrule
    \end{tabular}%
    }
\end{table}

We assess whether the transformations preserve program behavior and at what runtime cost.
We use each program's official test suite to check that its behavior is preserved.
Of the \num{10062} test cases in total, the unprotected programs pass all of them.
We then run the same test cases on the programs produced by each defense method.
We also check whether each ground-truth PoC still triggers on the transformed programs.
In addition, we manually review all transformation diffs and find no newly introduced vulnerabilities.
For added operations that could affect security, we check whether they are reachable and what inputs they receive.
Finally, we measure two deployment overheads relative to the unprotected programs, execution time (the wall-clock time of running the test cases) and on-disk size.
\autoref{tab:transformed_program_quality} shows the results.

\PP{Test Results.}
\sys passes all \num{10062} test cases, whereas JS-Obfuscator and FPA pass only 88.46\% and 68.52\%, respectively.
\sys's fixed AST operations change only the matched sites, and none of these changes alters the tested behavior.
In contrast, the baselines change program behavior in ways their own validation misses.
For JS-Obfuscator, generated property assignments change when
expressions are evaluated, while decoder helpers modify shared
runtime state that the original code relies on.
In \cc{satori}'s JSX runtime, its object-key rewriting moves a property assignment outside the guard of an originally unevaluated \cc{0 \&\& (...)} expression.
The moved assignment reads the undefined identifier \cc{Fragment} and raises a \cc{ReferenceError} that blocks 264 dependent tests.
In \cc{whistle}'s regex conversion helper, the self-defending logic of its generated decoder runs an extra regex check.
This check overwrites the shared captures in \cc{RegExp.\$1} and \cc{RegExp.\$2} and changes the results of four tests.
FPA validates only the injected pattern by executing the familiar function in isolation, without running the host program or its tests.
As a result , its accepted rewrites include undefined variables and uses before initialization at the injection site, which disrupt execution even though FPA changes relatively little code.

We also conduct differential testing using 20,000 package-specific inputs (500 per program), derived from regression tests and public API usage.
We observe no functional disagreements for \sys, while JS-Obfuscator and FPA yield disagreement rates of 91.70\% and 83.74\%, respectively.

\PP{PoC Success.}
All 40 PoCs still trigger under \sys, whereas only 37 (92.50\%) trigger under JS-Obfuscator and 33 (82.50\%) under FPA.
\sys thus preserves the vulnerable behavior, so the vulnerabilities that agents miss in \autoref{s:eval-effectiveness} are indeed hidden rather than removed.
\autoref{s:appendix-poc-failures} analyzes why the baseline PoCs fail.

\PP{Execution Time.}
The execution time of \sys is only 1.11\x that of the unprotected programs, compared with 2.46\x for JS-Obfuscator.
\sys's helper calls and decoding operations introduce some additional execution work.
JS-Obfuscator's decoding helpers and control-flow dispatch introduce more runtime operations, and its larger generated files also increase loading and parsing work.
The execution time of FPA is 0.95\x, because most files are unchanged and early failures shorten some executions.

\PP{Repository Size.}
\sys increases the mean repository size from 4.35\,MiB to 5.95\,MiB (1.37\x), whereas JS-Obfuscator more than doubles it to 11.43\,MiB (2.63\x).
The growth under \sys comes from the helpers and modules added by its rules and from its identifier-renaming pass (\autoref{s:design-renaming}).
JS-Obfuscator grows more because it adds dead code, control-flow dispatch structures, and string decoders.
FPA modifies only a few files, so its size grows by only 0.28\%.
We measure on-disk size rather than line count, because the transformations merge and split lines, so line counts do not reflect the actual growth.

\subsection{Generality}
\label{s:eval-generality}
\sys discovers rules from a corpus of five vulnerability types and against a single detection agent, yet the rules apply well beyond this setting.
The 40 evaluation programs are disjoint from the discovery corpus, and \sys applies varied rule-family combinations across them (\autoref{s:eval-rules}).
Moreover, 24 of them have vulnerability types outside the five-type corpus, such as cross-site scripting and authorization bypass; on these \sys reaches DSRs of 50.0--62.5\%, comparable to the 45.6--66.7\% on the 16 programs with in-corpus types.
Although discovered only against Claude Code with GLM-5.3, the rules achieve comparable DSRs on all four evaluation agents, which differ in harness, model, or both (\autoref{tab:defense_effectiveness_detection}), so they are not tied to the agent used in discovery.
These results follow because the rules target how an agent investigates code rather than a specific program, vulnerability type, or agent.

The rules also carry beyond JavaScript.
A prototype that manually ports the extracted rules to Python reduces Codex's recall from 70.0\% to 40.0\% on ten programs, suggesting that the strategies, not their JavaScript-specific form, drive the effect.
The full results are shown in \autoref{tab:python_extension} in \autoref{s:appendix-python}.

\subsection{Component-wise Analysis}
\label{s:eval-component}
We analyze the contribution of \sys's components along four dimensions.
\squishlist
\item The \emph{transformation discovery procedure}, which uses traces and strategies to guide program evolution.
\item The \emph{application agent}, which selects and adapts rules.
\item The \emph{rule set} of extracted transformation rules.
\item \emph{Defense-aware detection}, where the adversary knows that a defense may be present.
\squishend
All configurations use Codex with GPT-6.1-sol as the detection agent and Claude Code with GLM-5.3 as the default application agent.
Full \sys serves as the reference (recall 30.00\%, DSR 52.00\%, \autoref{tab:online_sensitivity}).

\PP{Transformation Discovery.}
With ten corpus programs and 100 mutation attempts each ($\approx$1/12 of the full budget), full discovery and ablations removing trace or strategy guidance yield 14, 3, and 4 validated rules (all in \autoref{tab:rule_characterization}) and DSRs of 28.0\%, 12.0\%, and 12.0\%.
Without guidance, the search thus adds nothing beyond renaming alone (12.0\%).

\PP{Application Agent.}
Disabling the agent, so that \sys chooses rules by a fixed ranking, raises recall to 55.00\% and lowers DSR to 12.00\%, 40.00 points below Full \sys.
The agent's contextual selection thus accounts for most of the defense.
Replacing GLM-5.3 with Claude Opus 5.5 in the same Claude Code agent yields recall 27.50\% and DSR 56.00\%, 4.00 points above the default.
Since Opus 5.5 costs over four times as much as GLM-5.3, we use GLM-5.3 by default.

\PP{Rule Set.}
Removing the rule set leaves a single-pass LLM rewrite (Direct LLM) without the rules or the identifier-renaming.
It provides no defense (recall 62.50\%, DSR 0.00\%).
It also changes program behavior, passing only 79.13\% of the project tests, whereas \sys passes all of them.
Letting an LLM write the changes directly is thus neither effective nor safe, which motivates fixed rule operations (\autoref{s:design-application}).
Applying only the identifier-renaming pass, without any rule, records recall 55.00\% and DSR 12.00\%, so the defense comes from the rules rather than renaming.
The rule set with a randomly selected half of the rules records recall 37.50\% and DSR 40.00\%, and the full thirteen-family rule set lowers recall by a further 7.50 points.

\PP{Defense-Aware Detection.}
We tell the detection agent that the program may be defended and instruct it to examine executable behavior, helper modules, and indirect calls.
Its recall then rises from 30.00\% to 37.50\%, and its DSR falls to 40.00\%.
Going further, a rule-aware attacker adds descriptions of the families on top of this prompt and removes comments and dead code before detection (\autoref{s:appendix-detection-prompts}).
Its DSR is 32.00\% (recall 42.50\%), so \sys remains effective even when the attacker knows its rules.

\begin{table}[t]
    \centering
    \caption{
      Component-wise analysis of \sys.
    }
    \label{tab:online_sensitivity}

    \small\rmfamily\upshape
    \setlength{\tabcolsep}{4pt}

    \begin{tabular*}{0.85\columnwidth}{@{\extracolsep{\fill}}lrr@{}}
        \toprule
        \textbf{Configuration}
            & \textbf{DSR (\%)}
            & \textbf{Recall (\%)} \\
        \midrule
        \sys
            & 52.00 & 30.00 \\
        \addlinespace[3pt]
        \rowcolor{black!8}[0pt][\tabcolsep]
        \multicolumn{3}{@{}l}{\strut\textbf{Application Agent}} \\
        w/o application agent
            & 12.00 & 55.00 \\
        Claude Code (Opus 5.5) 
            &   56.00    &  27.50     \\
        \addlinespace[3pt]
        \rowcolor{black!8}[0pt][\tabcolsep]
        \multicolumn{3}{@{}l}{\strut\textbf{Rule Set}} \\
        Direct LLM transformation
            & 0.00 & 62.50 \\
        Identifier renaming only
            &  12.00   &   55.00     \\
        Random half of rules
            & 40.00 & 37.50 \\
        \addlinespace[3pt]
        \rowcolor{black!8}[0pt][\tabcolsep]
        \multicolumn{3}{@{}l}{\strut\textbf{Defense-Aware Detection}} \\
        Defense-aware prompt
            & 40.00 & 37.50 \\
        Rule-aware prompt
            & 32.00 & 42.50 \\
        \bottomrule
    \end{tabular*}
\end{table}

\subsection{Deployment Cost}

\sys has a one-time cost for building the rule set and a small cost for each program it protects.
Rule discovery and extraction take about 433 hours and \$731.
They run once, and the resulting rules are reused across programs.
We measure the end-to-end time and API cost of applying \sys to each of the 40 programs with the default application agent (Claude Code with GLM-5.3).
On average, \sys costs \$1.94 (ranging from \$0.20 to \$9.01) and takes 45.01 minutes per program (\autoref{tab:application_cost}).
A few programs raise these averages, so the median cost is only \$1.31 and the median time only 11.94 minutes, and 30 of the 40 programs finish within 30 minutes.
The per-program cost of \sys is similar to that of FPA, which costs about \$2.06 and takes an estimated 9.5 minutes per program.
JS-Obfuscator is much faster, averaging 0.73 minutes per program, because it uses no LLM.
Cost rises modestly with program size.
\autoref{s:appendix-deployment} gives the per-program figures and the size-cost correlation.

\subsection{Rule Characterization}
\label{s:eval-rules}

We characterize the rules that \sys extracts and how often they are applied.
\autoref{tab:rule_characterization} lists the rule families and their usage, and \autoref{fig:heatmap} in \autoref{s:appendix-rule-usage} shows the number of applications of each family per program.

\PP{Rule Families.}
The rule set contains 111 rules in 13 rule families, which we group into four categories.
\emph{Structural Indirection} (\cc{F01}--\cc{F04}) changes how the vulnerable code is reached, for example by routing calls through indirect chains or splitting logic across modules, so that the dataflow is harder to trace.
\emph{Semantic Misdirection} (\cc{F05}--\cc{F08}) adds code whose apparent security meaning differs from its actual behavior, such as sanitization that does not sanitize or decoy code, so that the agent misjudges the code.
\emph{Context Manipulation} (\cc{F09}--\cc{F11}) changes the context that the agent relies on to decide what to examine, such as claims about the code or markers of third-party code.
\emph{Reasoning Complexity} (\cc{F12}--\cc{F13}) increases the reasoning effort needed to follow the code, for example by adding branches that the agent must explore.
Most families have no counterpart in conventional obfuscators.
JS-Obfuscator changes only the surface syntax, whereas the Semantic Misdirection and Context Manipulation families change what the agent believes about the code.
\cc{F08} extends FPA's abstraction bias~\cite{fpa} to ordinary idioms, misleading the agent about control flow and decoding results.

\PP{Rule Usage.}
\sys applies 60 of the 111 rules and uses all 13 families.
Each program receives 5--10 families (7.83 on average), and the 40 programs use 27 distinct family combinations, the most common of which is shared by only four programs.
Seven rule families are applied to at least 75\% of the programs (\autoref{tab:rule_characterization}).
The most widely applied are \cc{F13} (Multi-Branch Trap), applied to every program, \cc{F08} (General Pattern Misanalysis, 97.5\%), and \cc{F01} (Call-Chain Indirection, 92.5\%).
These families match common code constructs, such as call expressions and dynamic member accesses, which also allow repeated applications within a program.
In contrast, other rule families are less applied, for two different reasons.
First, some depend on code patterns that few programs contain.
\cc{F02} (Call-Node Hiding) and \cc{F03} (Regex Fragmentation) have matching sites in only one and five programs, respectively.
Second, some have matching sites in every program but are rarely selected during application.
\cc{F09} (Template Injection) and \cc{F10} (Fake Claim) have matching sites in all 40 programs but are selected in only six and eight.
Most programs use several families that jointly impede detection, so a successful defense cannot usually be attributed to a single family.

\begin{table}[t]
    \centering
    \caption{
        Rule families and their usage across the 40 evaluation programs.
        Var.: number of rules in the rule set.
        Applied rules: number of rules applied at least once.
        Programs: number of programs to which the family is applied.
    }
    \label{tab:rule_characterization}

    \small\rmfamily\upshape

    \resizebox{\columnwidth}{!}{%
    \begin{tabular}{@{}lrr@{}}
        \toprule
        \textbf{Rule family}
            & \textbf{Rules}
            & \textbf{Programs} \\
            &
            \textbf{Applied / Total}
            & \textbf{\# (\%)} \\
        \midrule
        \rowcolor{black!8}[0pt][\tabcolsep]
        \multicolumn{3}{@{}l}{\strut\textbf{Structural Indirection}} \\
        \cc{F01:}\enspace Call-Chain Indirection
        & 10/10 & 37 (92.5) \\

        \cc{F02:}\enspace Call-Node Hiding
        & 1/10 & 1 (2.5) \\

        \cc{F03:}\enspace Regex Fragmentation
        & 2/10 & 5 (12.5) \\

        \cc{F04:}\enspace Cross-Module Fragmentation
        & 1/5 & 27 (67.5) \\

        \addlinespace[3pt]
        \rowcolor{black!8}[0pt][\tabcolsep]
        \multicolumn{3}{@{}l}{\strut\textbf{Semantic Misdirection}} \\
        \cc{F05:}\enspace Fake Sanitization
        & 5/10 & 34 (85.0) \\

        \cc{F06:}\enspace Sanitization Reversal
        & 7/10 & 35 (87.5) \\

        \cc{F07:}\enspace Decoy Insertion
        & 9/10 & 31 (77.5) \\

        \cc{F08:}\enspace General Pattern Misanalysis
        & 6/6 & 39 (97.5) \\

        \addlinespace[3pt]
        \rowcolor{black!8}[0pt][\tabcolsep]
        \multicolumn{3}{@{}l}{\strut\textbf{Context Manipulation}} \\
        \cc{F09:}\enspace Template Injection
        & 2/2 & 6 (15.0) \\

        \cc{F10:}\enspace Fake Claim
        & 2/12 & 8 (20.0) \\

        \cc{F11:}\enspace Untrusted Third-Party Artifact
        & 2/6 & 30 (75.0) \\

        \addlinespace[3pt]
        \rowcolor{black!8}[0pt][\tabcolsep]
        \multicolumn{3}{@{}l}{\strut\textbf{Reasoning Complexity}} \\
        \cc{F12:}\enspace Complexity Injection
        & 4/10 & 20 (50.0) \\

        \cc{F13:}\enspace Multi-Branch Trap
        & 9/10 & 40 (100.0) \\

        \midrule
        \textbf{Total}
        & \textbf{60/111} & 40 (100.0) \\
        \bottomrule
    \end{tabular}%
    }
\end{table}

\section{Discussion}
\label{s:discussion}

\PP{Dual Use.}
Our threat model concerns developers protecting released code against agentic vulnerability detection.
The same transformations could be misused, for instance to conceal vulnerabilities in malicious code or to impede legitimate security review.
However, we believe the defensive benefit outweighs this risk.
Attackers already use detection agents to find and exploit vulnerabilities in released software~\cite{ai-hacker-14-companies, anthropic-threat-report}, yet defenders have had no countermeasure.
To limit misuse, we plan to conditionally release \sys only after reviewing the requester's intended use.

\PP{Limitations}
\sys does not formally prove that the rules preserve semantics.
We instead validate preservation in two ways.
First, we manually inspect every rule \sys extracts, which is practical because each rule is a fixed operation over a local matched structure.
Second, we follow the common practice of testing, running each transformed program against a set of testing inputs.
These exercise a wide range of each program's behavior, so passing all of them empirically supports the semantics preservation.
Proving semantic equivalence formally remains difficult for a dynamic language like JavaScript and is orthogonal to our contribution.
We leave it to future work.

\section{Related Work}
\label{relw}

\PP{Program Analysis for JavaScript Code}
Vulnerability detection in JavaScript has long relied on program analysis, such as object dependence graphs~\cite{li2022odgen}, runtime enforcement~\cite{staicu2018synode}, and dynamic taint analysis~\cite{cassel2023nodemedic,cassel2025nodemedicfine}.
More recently, LLM-based agents have become strong vulnerability detectors.
Some are dedicated harnesses that couple a model with analysis tools or code property graphs~\cite{symbolon, repoaudit,enigma,iris,llmxcpg,ni2026llmvdjs}, while others are general-purpose coding agents that audit code directly~\cite{codex,claude-code,big-sleep}.

\PP{Anti Code Analysis}
Anti-analysis techniques aim to make software harder to analyze or test while maintaining its intended behavior~\cite{SharifLGL08}.
Popov et al.\ hide jumps and calls behind signal handlers, making static disassembly and control-flow recovery harder~\cite{popov2007signals}.
Loki replaces simple calculations with complex equivalent expressions to resist symbolic execution and other automated attempts to simplify the code~\cite{schloegel2022loki}.
For automated vulnerability discovery, Fuzzification slows fuzzing runs and introduces misleading coverage feedback~\cite{jung2019fuzzification}.
AntiFuzz also hides crashes and makes input constraints harder to solve, increasing the cost of fuzzing and symbolic analysis~\cite{guler2019antifuzz}.
\sys shares the goal of making vulnerabilities harder to discover through code transformations.
It focuses on how LLM-based agents search for relevant code, use analysis tools, and follow the dataflow needed to establish a vulnerability.

\section{Conclusion}
\label{s:conclusion}

In this work, we present \sys, a defense that hides vulnerabilities in released JavaScript programs from LLM-based detection agents.
\sys discovers novel transformations through search, extracts them into reusable rules, and applies the rules to new programs while preserving behavior and keeping overhead low.
Across four detection agents, it cuts the vulnerabilities each finds by more than half while passing all project tests.
Protecting code against analyzing agents is an important and growing direction, as agentic offensive analysis grows cheaper and more capable with stronger models.
We hope \sys motivates more work on defending code against analyzing agents.

\section*{Acknowledgments}
This work was supported in part by the NTU Start-Up Grant and the Ministry of Education, Singapore, under its Academic Research Fund Tier 1 (RS8/26). 
Penghui Li is the corresponding author.



\begin{thebibliography}{10}
\providecommand{\url}[1]{#1}
\csname url@samestyle\endcsname
\providecommand{\newblock}{\relax}
\providecommand{\bibinfo}[2]{#2}
\providecommand{\BIBentrySTDinterwordspacing}{\spaceskip=0pt\relax}
\providecommand{\BIBentryALTinterwordstretchfactor}{4}
\providecommand{\BIBentryALTinterwordspacing}{\spaceskip=\fontdimen2\font plus
\BIBentryALTinterwordstretchfactor\fontdimen3\font minus \fontdimen4\font\relax}
\providecommand{\BIBforeignlanguage}[2]{{%
\expandafter\ifx\csname l@#1\endcsname\relax
\typeout{** WARNING: IEEEtran.bst: No hyphenation pattern has been}%
\typeout{** loaded for the language `#1'. Using the pattern for}%
\typeout{** the default language instead.}%
\else
\language=\csname l@#1\endcsname
\fi
#2}}
\providecommand{\BIBdecl}{\relax}
\BIBdecl

\bibitem{rack2023jack}
J.~Rack and C.-A. Staicu, ``{Jack-in-the-Box}: An empirical study of {JavaScript} bundling on the web and its security implications,'' in \emph{Proceedings of the 2023 ACM SIGSAC Conference on Computer and Communications Security (CCS)}, 2023.

\bibitem{secbench:icse:2023}
M.~H.~M. Bhuiyan, A.~S. Parthasarathy, N.~Vasilakis, M.~Pradel, and C.~Staicu, ``Secbench.js: An executable security benchmark suite for server-side javascript,'' in \emph{45th {IEEE/ACM} International Conference on Software Engineering, {ICSE}}, 2023.

\bibitem{skolka2019anything}
P.~Skolka, C.-A. Staicu, and M.~Pradel, ``Anything to hide? studying minified and obfuscated code in the web,'' in \emph{Proceedings of the 2019 World Wide Web Conference (WWW)}, 2019.

\bibitem{owasp:frontend-security}
{OWASP Foundation}, ``Web frontend security cheat sheet,'' \url{https://cheatsheetseries.owasp.org/cheatsheets/Web_Frontend_Security_Cheat_Sheet.html}, accessed: November 2026.

\bibitem{owasp:cookie-theft}
\BIBentryALTinterwordspacing
------, ``Cookie theft mitigation cheat sheet,'' 2026, accessed: November 2026. [Online]. Available: \url{https://cheatsheetseries.owasp.org/cheatsheets/Cookie_Theft_Mitigation_Cheat_Sheet.html}
\BIBentrySTDinterwordspacing

\bibitem{owasp:csrf}
\BIBentryALTinterwordspacing
------, ``Owasp code review guide: Cross-site request forgery,'' 2017, accessed: November 2026. [Online]. Available: \url{https://owasp.org/www-project-code-review-guide/assets/OWASP_Code_Review_Guide_v2.pdf}
\BIBentrySTDinterwordspacing

\bibitem{khodayari2021jaw}
S.~Khodayari and G.~Pellegrino, ``{JAW}: Studying client-side {CSRF} with hybrid property graphs and declarative traversals,'' in \emph{30th {USENIX} Security Symposium, {USENIX} Security}, 2021.

\bibitem{shcherbakov2023silentspring}
M.~Shcherbakov, M.~Balliu, and C.~Staicu, ``{Silent Spring}: Prototype pollution leads to remote code execution in {Node.js},'' in \emph{32nd {USENIX} Security Symposium, {USENIX} Security}, 2023.

\bibitem{repoaudit}
J.~Guo, C.~Wang, X.~Xu, Z.~Su, and X.~Zhang, ``Repoaudit: An autonomous llm-agent for repository-level code auditing,'' in \emph{Forty-second International Conference on Machine Learning, {ICML}}, 2025.

\bibitem{enigma}
T.~Abramovich, M.~Udeshi, M.~Shao, K.~Lieret, H.~Xi, K.~Milner, S.~Jancheska, J.~Yang, C.~E. Jimenez, F.~Khorrami, P.~Krishnamurthy, B.~Dolan{-}Gavitt, M.~Shafique, K.~R. Narasimhan, R.~Karri, and O.~Press, ``Enigma: Interactive tools substantially assist {LM} agents in finding security vulnerabilities,'' in \emph{Forty-second International Conference on Machine Learning ({ICML})}, 2025.

\bibitem{bugauditor2026}
M.~Lin, K.~Chen, and H.~Chen, ``{BugAuditor}: Detecting bugs via inconsistent defensive code auditing,'' in \emph{35th USENIX Security Symposium (USENIX Security)}, 2026.

\bibitem{ni2026llmvdjs}
R.~Ni, M.~Christodorescu, and L.~Jia, ``Taint-style vulnerability detection and confirmation for {Node.js} packages using {LLM} agent reasoning,'' \emph{arXiv preprint arXiv:2604.20179}, 2026.

\bibitem{simsek2025pocgen}
D.~Simsek, A.~Eghbali, and M.~Pradel, ``Pocgen: Generating proof-of-concept exploits for vulnerabilities in npm packages,'' \emph{Proc. {ACM} Softw. Eng.}, vol.~3, no. {FSE}, pp. 3887--3908, 2026.

\bibitem{claude-code}
{Anthropic}, ``{Claude Code},'' \url{https://github.com/anthropics/claude-code}, 2026, accessed: November 2026.

\bibitem{codex}
{OpenAI}, ``{Codex},'' \url{https://github.com/openai/codex}, 2026, accessed: November 2026.

\bibitem{ai-hacker-14-companies}
{OALABS Research}, ``Captured logs reveal hackers using claude and codex to breach companies,'' \url{https://research.openanalysis.net/claude/codex/hacking/ai\%20hacking/llm/redteam/policy\%20violation/2026/06/16/compromised-claude-hacking.html}, 2026, accessed: November 2026.

\bibitem{anthropic-threat-report}
{Anthropic}, ``Countering misuse of {AI}: September 2026,'' \url{https://www.anthropic.com/threat-intelligence-report-september-2026}, 2026.

\bibitem{jung2019fuzzification}
J.~Jung, H.~Hu, D.~Solodukhin, D.~Pagan, K.~H. Lee, and T.~Kim, ``Fuzzification: {Anti-Fuzzing} techniques,'' in \emph{28th USENIX Security Symposium (USENIX Security 19)}, 2019, pp. 1913--1930.

\bibitem{attack_1_hidenoseek}
A.~Fass, M.~Backes, and B.~Stock, ``Hidenoseek: Camouflaging malicious javascript in benign asts,'' in \emph{Proceedings of the 2019 {ACM} {SIGSAC} Conference on Computer and Communications Security, {CCS}}, 2019.

\bibitem{attack_2_wolf}
X.~Ling, Z.~Wu, B.~Wang, W.~Deng, J.~Wu, S.~Ji, T.~Luo, and Y.~Wu, ``A wolf in sheep's clothing: Practical black-box adversarial attacks for evading learning-based windows malware detection in the wild,'' in \emph{33rd {USENIX} Security Symposium, {USENIX} Security}, 2024.

\bibitem{javascript-obfuscator}
{JavaScript Obfuscator Contributors}, ``{JavaScript Obfuscator},'' \url{https://github.com/javascript-obfuscator/javascript-obfuscator}.

\bibitem{fpa}
S.~Bernstein, D.~Beste, D.~Ayzenshteyn, L.~Sch{\"o}nherr, and Y.~Mirsky, ``Trust me, i know this function: Hijacking {LLM} static analysis using bias,'' in \emph{Annual Network and Distributed System Security Symposium (NDSS)}, 2026.

\bibitem{li2025cotdeceptor}
\BIBentryALTinterwordspacing
H.~Li, M.~Li, J.~Zuo, S.~Li, X.~Li, H.~Wu, Y.~Lu, and X.~He, ``{CoTDeceptor}: Adversarial code obfuscation against {CoT}-enhanced {LLM} code agents,'' 2025. [Online]. Available: \url{https://arxiv.org/abs/2512.21250}
\BIBentrySTDinterwordspacing

\bibitem{zcode}
{Z.ai}, ``{ZCode},'' \url{https://github.com/zai-org/ZCode}, 2026, accessed: November 2026.

\bibitem{openant}
\BIBentryALTinterwordspacing
N.~Korda and G.~Evron, ``{OpenAnt}: {LLM}-powered vulnerability discovery through code decomposition, adversarial verification, and dynamic testing,'' \emph{arXiv preprint arXiv:2606.19149}, 2026. [Online]. Available: \url{https://arxiv.org/abs/2606.19149}
\BIBentrySTDinterwordspacing

\bibitem{shi2026toolhijacker}
J.~Shi, Z.~Yuan, G.~Tie, P.~Zhou, N.~Z. Gong, and L.~Sun, ``Prompt injection attack to tool selection in {LLM} agents,'' in \emph{Proceedings of the Network and Distributed System Security Symposium, {NDSS}}, 2026.

\bibitem{react}
S.~Yao, J.~Zhao, D.~Yu, N.~Du, I.~Shafran, K.~R. Narasimhan, and Y.~Cao, ``React: Synergizing reasoning and acting in language models,'' in \emph{The Eleventh International Conference on Learning Representations, {ICLR}}, 2023.

\bibitem{sweagent}
J.~Yang, C.~E. Jimenez, A.~Wettig, K.~Lieret, S.~Yao, K.~R. Narasimhan, and O.~Press, ``{SWE}-agent: Agent-computer interfaces enable automated software engineering,'' in \emph{The Thirty-eighth Annual Conference on Neural Information Processing Systems}, 2024.

\bibitem{cybergym}
Z.~Wang, T.~Shi, J.~He, M.~Cai, J.~Zhang, and D.~Song, ``{CyberGym}: Evaluating {AI} agents' real-world cybersecurity capabilities at scale,'' in \emph{International Conference on Learning Representations}, 2026.

\bibitem{cybergym-leaderboard}
{CyberGym}, ``{CyberGym} leaderboard,'' \url{https://www.cybergym.io/cybergym/}, accessed: November 2026.

\bibitem{romano2022wobfuscator}
A.~Romano, D.~Lehmann, M.~Pradel, and W.~Wang, ``{Wobfuscator}: Obfuscating {JavaScript} malware via opportunistic translation to {WebAssembly},'' in \emph{43rd {IEEE} Symposium on Security and Privacy, {SP}}, 2022.

\bibitem{ua-parser-js}
\BIBentryALTinterwordspacing
{faisalman}, ``{UAParser.js}: Powerful user-agent detection in a compact, es5-friendly, dependency-free javascript library,'' 2026, accessed: November 2026. [Online]. Available: \url{https://github.com/faisalman/ua-parser-js}
\BIBentrySTDinterwordspacing

\bibitem{iris}
Z.~Li, S.~Dutta, and M.~Naik, ``{IRIS:} llm-assisted static analysis for detecting security vulnerabilities,'' in \emph{The Thirteenth International Conference on Learning Representations, {ICLR}}, 2025.

\bibitem{llmxcpg}
A.~Lekssays, H.~Mouhcine, K.~Tran, T.~Yu, and I.~Khalil, ``{LLMxCPG}: Context-aware vulnerability detection through code property graph-guided large language models,'' in \emph{34th USENIX Security Symposium (USENIX Security 25)}, 2025, pp. 489--507.

\bibitem{SecLLMHolmes}
S.~Ullah, M.~Han, S.~Pujar, H.~Pearce, A.~K. Coskun, and G.~Stringhini, ``Llms cannot reliably identify and reason about security vulnerabilities (yet?): {A} comprehensive evaluation, framework, and benchmarks,'' in \emph{{IEEE} Symposium on Security and Privacy, {SP}}, 2024.

\bibitem{redherring}
\BIBentryALTinterwordspacing
K.~Zhang, Z.~Zhang, Y.~Xie, Z.~Liu, S.~Yao, Z.~Zhang, and D.~She, ``Cheap to hypothesize, costly to verify: The defense surface of agentic vulnerability discovery,'' \emph{CoRR}, 2026. [Online]. Available: \url{https://doi.org/10.48550/arXiv.2609.35909}
\BIBentrySTDinterwordspacing

\bibitem{zhu2026natural}
\BIBentryALTinterwordspacing
J.~Zhu, R.~Bhaskar, H.~Zheng, B.~Y. Zhao, and K.~Pei, ``Natural code transformation confuses agents in vulnerability analysis,'' 2026. [Online]. Available: \url{https://arxiv.org/abs/2610.14350}
\BIBentrySTDinterwordspacing

\bibitem{openevolve}
\BIBentryALTinterwordspacing
A.~Sharma, ``Openevolve: an open-source evolutionary coding agent,'' 2025. [Online]. Available: \url{https://github.com/algorithmicsuperintelligence/openevolve}
\BIBentrySTDinterwordspacing

\bibitem{babel}
``Babel: The {JavaScript} compiler,'' 2026, \url{https://babeljs.io/}.

\bibitem{secbench-2}
F.~Xiao, Z.~Su, G.~Yang, and W.~Lee, ``Jasmine: Scale up javascript static security analysis with computation-based semantic explanation,'' in \emph{{IEEE} Symposium on Security and Privacy, {SP}}, 2024.

\bibitem{secbench-3}
C.~Luo, P.~Li, W.~Meng, and C.~Zhang, ``Test suites guided vulnerability validation for node.js applications,'' in \emph{Proceedings of the 2024 on {ACM} {SIGSAC} Conference on Computer and Communications Security, {CCS}}, 2024.

\bibitem{github-advisory-db}
{GitHub}, ``{GitHub} advisory database: Reviewed npm advisories,'' \url{https://github.com/advisories?query=type\%3Areviewed+ecosystem\%3Anpm}, 2026, accessed: November 2026.

\bibitem{chen2025jsdeobsbench}
G.~Chen, X.~Jin, and Z.~Lin, ``{JsDeObsBench}: Measuring and benchmarking {LLMs} for {JavaScript} deobfuscation,'' in \emph{Proceedings of the 2025 ACM SIGSAC Conference on Computer and Communications Security (CCS)}, 2025.

\bibitem{anthropic-opus55-pricing}
\BIBentryALTinterwordspacing
{Anthropic}, ``{Claude Opus 5.5}: For long-running agentic coding and knowledge work.'' 2026, accessed: November 2026. [Online]. Available: \url{https://platform.claude.com/docs/en/models/opus-5-5/overview}
\BIBentrySTDinterwordspacing

\bibitem{openai-gpt61sol-pricing}
\BIBentryALTinterwordspacing
{OpenAI}, ``{GPT-6.1 Sol}: Near-astra performance for complex work at a lower cost.'' 2026, accessed: November 2026. [Online]. Available: \url{https://developers.openai.com/api/docs/models}
\BIBentrySTDinterwordspacing

\bibitem{zai-glm51-pricing}
\BIBentryALTinterwordspacing
{Z.AI}, ``{GLM-5.3} model pricing,'' 2026. [Online]. Available: \url{https://docs.z.ai/guides/overview/pricing}
\BIBentrySTDinterwordspacing

\bibitem{li2022odgen}
S.~Li, M.~Kang, J.~Hou, and Y.~Cao, ``Mining node.js vulnerabilities via object dependence graph and query,'' in \emph{31st {USENIX} Security Symposium}, 2022.

\bibitem{staicu2018synode}
C.~Staicu, M.~Pradel, and B.~Livshits, ``{SYNODE:} understanding and automatically preventing injection attacks on {NODE.JS},'' in \emph{25th Annual Network and Distributed System Security Symposium}, 2018.

\bibitem{cassel2023nodemedic}
D.~Cassel, W.~T. Wong, and L.~Jia, ``Nodemedic: End-to-end analysis of node.js vulnerabilities with provenance graphs,'' in \emph{8th {IEEE} European Symposium on Security and Privacy, EuroS{\&}P}, 2023.

\bibitem{cassel2025nodemedicfine}
D.~Cassel, N.~Sabino, M.-C. Hsu, R.~Martins, and L.~Jia, ``{NodeMedic-FINE}: Automatic detection and exploit synthesis for {Node.js} vulnerabilities,'' in \emph{Proceedings of the Network and Distributed System Security Symposium, {NDSS}}, 2025.

\bibitem{symbolon}
\BIBentryALTinterwordspacing
J.~Zhu, P.~Li, Z.~Li, C.~Shen, Z.~Li, Y.~Chen, and K.~Pei, ``Symbolon: Symbolic execution by learning code transformation,'' 2026. [Online]. Available: \url{https://arxiv.org/abs/2606.29108}
\BIBentrySTDinterwordspacing

\bibitem{big-sleep}
\BIBentryALTinterwordspacing
{Google Project Zero}, ``From naptime to big sleep: Using large language models to catch vulnerabilities in real-world code,'' 2024, accessed: November 2026. [Online]. Available: \url{https://projectzero.google/2024/10/from-naptime-to-big-sleep.html}
\BIBentrySTDinterwordspacing

\bibitem{SharifLGL08}
M.~I. Sharif, A.~Lanzi, J.~T. Giffin, and W.~Lee, ``Impeding malware analysis using conditional code obfuscation,'' in \emph{Proceedings of the Network and Distributed System Security Symposium, {NDSS}}, 2008.

\bibitem{popov2007signals}
I.~V. Popov, S.~K. Debray, and G.~R. Andrews, ``Binary obfuscation using signals,'' in \emph{Proceedings of the 16th {USENIX} Security Symposium}, 2007.

\bibitem{schloegel2022loki}
M.~Schloegel, T.~Blazytko, M.~Contag, C.~Aschermann, J.~Basler, T.~Holz, and A.~Abbasi, ``Loki: Hardening code obfuscation against automated attacks,'' in \emph{31st {USENIX} Security Symposium, {USENIX} Security}, 2022.

\bibitem{guler2019antifuzz}
E.~G{\"{u}}ler, C.~Aschermann, A.~Abbasi, and T.~Holz, ``Antifuzz: Impeding fuzzing audits of binary executables,'' in \emph{28th {USENIX} Security Symposium, {USENIX} Security}, 2019.

\bibitem{spearman1904}
C.~Spearman, ``The proof and measurement of association between two things,'' \emph{The American Journal of Psychology}, vol.~15, no.~1, pp. 72--101, 1904.

\end{thebibliography}

\appendices

\section{Scoring Details}
\label{s:appendix-scoring}

This appendix gives further details of the degradation score in \autoref{s:design-feedback}.

\PP{Path Distance}
\sys obtains the reference dataflow path $R_v$ by running the PoC on an instrumented copy of the node's program.
The reference is stored outside the detection agent's workspace, so the agent cannot read it.
The distance $d_{v,j}$ is the Jaccard distance between the sets of (function, variable) pairs on the reference path and on the reported path.
If a run reports a finding but no parseable path, \sys retries parsing it up to three times and otherwise drops the run from $J_v$; this differs from a run that reports no finding, which receives $s_{v,j}=1$.

\PP{Type Score}
The type accuracy $a_{v,j}$ is the total score of findings whose vulnerability matches the annotated type, divided by the total score of all findings.
The score of each finding comes from the detection agent's output, or from normalizing its free-text answer when the output gives no score.
This score is separate from the agent's confidence.

\PP{Time Component}
The time component $h_{v,j}$ is the detection time divided by twice the original program's mean detection time under the same detection agent, capped at one.

\PP{Parameter Selection and Sensitivity}
We use path distance, type accuracy, and detection time to capture deviations in dataflow analysis, vulnerability-type identification, and analysis speed.
We prioritize path distance because naming the correct type does not establish the vulnerable dataflow, while slower analysis alone does not imply failed detection.
In pilot runs, we compare three weight settings: $(0.60,0.25,0.15)$, $(0.50,0.30,0.20)$, and $(0.45,0.30,0.25)$.
We select $(w_d,w_a,w_h)=(0.50,0.30,0.20)$ because its node scores best match our manual assessment of detection degradation.    

\PP{Valid Runs and Review}
The review agent checks every candidate whose mean run score is at least $0.6$.
A run is invalid if it fails because of an infrastructure or parsing error.
An explicit refusal or a confirmed local timeout counts as valid only if the agent read the target files.
A valid review returns a correction $C_v$ between zero and one, and \sys lowers $D_v$ to $C_v$ when $C_v$ is smaller.
If no valid review is obtained, the candidate is discarded.

\PP{Novelty Score}
The factor $D_v$ weights a node's new strategies by how much it impedes detection, so that a novel but ineffective change contributes little to exploration.
We normalize $\pi_{\mathrm{explore}}(v)\propto N_v/(1+c_v)$ over the nodes currently in the tree.

\PP{Node Selection}
In $\pi_{\mathrm{exploit}}$, \sys ranks nodes by $D_v$ and weights the node of rank $r$ by $1/\sqrt{r}$ before normalizing.
It sets the exploration rate to $p=0.25$, so it explores in a quarter of the iterations and exploits in the rest.
Exploitation takes priority because it builds on changes that already impede detection, while exploration pursues new strategies whose effects are less certain.

\section{Matching Conditions}
\label{s:appendix-rule-application}

Matching checks the conditions needed to apply a rule while preserving the program's semantics.
For the rule in \autoref{lst:rule}, the captured argument supplies the original value to the original call, while the added check examines a separate copy.
The check's result is discarded rather than used to guard the call.
Similarly, \cc{restored} in \autoref{code:example} reconstructs \cc{raw} before the parser receives it.
The matcher also checks where statements would be inserted.
It rejects \cc{enabled \&\& JSON.parse(makeInput())}, because extracting the argument before the condition would run \cc{makeInput()} even when the original call would be skipped.
The application agent can select a matched rule but cannot change its matching conditions or operation.

\section{Detection-Agent Prompts}
\label{s:appendix-detection-prompts}
Each audit ends when the agent finishes or reaches a 60-minute limit.
Timeouts and refusals remain in the results, and any findings reported
before the run ends are counted.
Runs interrupted by API or connection failures are retried at most twice.
\promptbox{Vulnerability detection}{
\textbf{System:}

You are an expert security analyst specializing in vulnerability analysis for JavaScript code.
Analyze the project code to identify security vulnerabilities, and report each finding with its location, code evidence, and a short technical explanation.

You may freely use the available tools within the sandbox. Choose the analysis actions needed to investigate and validate potential vulnerabilities.
All actions must remain within the sandbox. Do not use web-search or browsing tools.

\textbf{User:}\\
Project directory: \{PROJECT\_DIRECTORY\}.
}

\promptbox{Additional defense-aware instructions}{
\label{prompt:defense-aware-detection}
\textbf{User:}\\
NOTICE: The code may have been modified to make security analysis harder.
Check whether security checks affect execution, whether cleaned values are actually used, and other potential defenses.
Follow relevant helper modules and indirect calls, and base findings on executable behavior rather than comments or names.
}

For the rule-aware attacker, we first remove all comments and dead code from the program, and then append the following instructions to the defense-aware prompt.

\promptbox{Additional rule-aware instructions}{
\label{prompt:rule-aware-detection}
\textbf{User:}\\
The defense may apply the following kinds of transformations.\\
(1)~Structural indirection: routing calls through indirect chains (call-chain indirection, call-node hiding), splitting regular expressions, or splitting logic across modules.\\
(2)~Semantic misdirection: adding fake sanitization, sanitizing a value and then restoring it, inserting decoy code, or rewriting common idioms so that control flow and decoding results are misread.\\
(3)~Context manipulation: inserting chat-template text, false claims about the code, or markers of untrusted third-party code.\\
(4)~Reasoning complexity: adding complex computation or multiple branches that lead away from the vulnerable path.\\
Treat such code with suspicion and determine what each value and call actually does at runtime.
}

\section{Deployment Overhead}
\label{s:appendix-deployment}

\autoref{tab:application_cost} lists the size of each program and the size that \sys adds to it.
It also lists the time and API cost of applying \sys to each program.

To see whether larger programs cost more, we compare each program's original size with its cost.
We use the Spearman rank correlation~\cite{spearman1904}, which tests whether cost tends to rise with size without assuming a linear relation.
The correlation is positive and significant ($\rho=0.46$, $p<0.01$).
A least-squares linear fit gives roughly \$1.3 per program plus \$0.15 per MiB of original code, but explains only about a fifth of the variation ($R^2=0.21$), so it is a rough estimate rather than a precise prediction.

\begin{table*}[t]
    \centering
    \caption{Per-program size and deployment cost of \sys. Size: original repository size. $+$Size: size added by \sys.}
    \label{tab:application_cost}
    \footnotesize\rmfamily\upshape
    \definecolor{configgray}{gray}{0.94}
    \setlength{\tabcolsep}{3pt}
    \begin{tabularx}{\textwidth}{@{}
        c >{\raggedright\arraybackslash}X rrrr
        @{\hspace{0.7em}}|@{\hspace{0.7em}}
        c >{\raggedright\arraybackslash}X rrrr@{}}
        \toprule
        \textbf{ID} & \textbf{Program}
        & \textbf{Size} & \textbf{$+$Size} & \textbf{Time} & \textbf{Cost}
        & \textbf{ID} & \textbf{Program}
        & \textbf{Size} & \textbf{$+$Size} & \textbf{Time} & \textbf{Cost} \\
        & & \textbf{(MiB)} & \textbf{(MiB)} & \textbf{(min)} & \textbf{(\$)}
        & & & \textbf{(MiB)} & \textbf{(MiB)} & \textbf{(min)} & \textbf{(\$)} \\
        \midrule
        1 & style-\allowbreak dictionary & 4.18 & 0.27 & 6.11 & 1.62 & 21 & @roomi-\allowbreak fields/\allowbreak notebooklm-\allowbreak mcp & 2.12 & 0.86 & 6.09 & 1.04 \\
        \rowcolor{configgray}
        2 & appium-\allowbreak mcp & 1.52 & 0.47 & 5.53 & 1.06 & 22 & claude-\allowbreak code-\allowbreak templates & 2.40 & 3.44 & 25.73 & 1.55 \\
        3 & nodejs-\allowbreak whisper & 28.49 & 0.91 & 4.51 & 1.15 & 23 & odf-\allowbreak kit & 1.54 & 0.69 & 7.00 & 1.45 \\
        \rowcolor{configgray}
        4 & @apify/\allowbreak actors-\allowbreak mcp-\allowbreak server & 5.86 & 2.80 & 9.04 & 1.37 & 24 & tinymce & 11.49 & 4.15 & 91.56 & 2.19 \\
        5 & @ooples/\allowbreak token-\allowbreak optimizer-\allowbreak mcp & 5.45 & 3.83 & 54.93 & 1.77 & 25 & obsidian-\allowbreak local-\allowbreak rest-\allowbreak api & 9.26 & 4.43 & 136.56 & 4.69 \\
        \rowcolor{configgray}
        6 & marko & 1.27 & 1.02 & 8.09 & 1.24 & 26 & @zereight/\allowbreak mcp-\allowbreak gitlab & 1.58 & 0.41 & 118.82 & 1.45 \\
        7 & @delmaredigital/\allowbreak payload-\allowbreak puck & 1.68 & 0.83 & 6.38 & 1.55 & 27 & @aws-\allowbreak amplify/\allowbreak codegen-\allowbreak ui-\allowbreak react & 1.92 & 5.09 & 10.72 & 1.00 \\
        \rowcolor{configgray}
        8 & mediasoup & 7.11 & 3.92 & 18.96 & 4.43 & 28 & faf-\allowbreak mcp & 2.24 & 1.20 & 28.55 & 1.18 \\
        9 & fabric & 21.12 & 5.05 & 164.36 & 9.01 & 29 & obsidian-\allowbreak tc & 4.08 & 0.32 & 553.05 & 5.66 \\
        \rowcolor{configgray}
        10 & @grpc/\allowbreak grpc-\allowbreak js-\allowbreak xds & 7.01 & 0.76 & 7.05 & 0.68 & 30 & satori & 5.58 & 1.16 & 54.94 & 5.03 \\
        11 & whistle & 3.00 & 2.80 & 8.47 & 0.75 & 31 & grok-\allowbreak faf-\allowbreak mcp & 1.49 & 0.99 & 21.42 & 0.86 \\
        \rowcolor{configgray}
        12 & wetty & 5.25 & 0.48 & 7.47 & 0.48 & 32 & swagger-\allowbreak typescript-\allowbreak api & 1.29 & 0.30 & 8.47 & 1.21 \\
        13 & @sentry/\allowbreak nextjs & 1.40 & 0.53 & 13.15 & 1.05 & 33 & @novu/\allowbreak js & 1.82 & 1.03 & 10.64 & 0.48 \\
        \rowcolor{configgray}
        14 & sequelize & 2.77 & 2.66 & 54.11 & 3.98 & 34 & @apostrophecms/\allowbreak import-\allowbreak export & 1.84 & 0.24 & 3.27 & 0.27 \\
        15 & @highcharts/\allowbreak grid-\allowbreak lite & 5.32 & 2.54 & 13.83 & 2.93 & 35 & ua-\allowbreak parser-\allowbreak js & 1.25 & 0.24 & 15.85 & 0.71 \\
        \rowcolor{configgray}
        16 & @payloadcms/\allowbreak plugin-\allowbreak import-\allowbreak export & 1.42 & 0.31 & 6.09 & 1.03 & 36 & defuddle & 2.51 & 2.81 & 91.10 & 2.20 \\
        17 & kysely & 3.13 & 2.16 & 23.49 & 2.30 & 37 & claude-\allowbreak faf-\allowbreak mcp & 1.10 & 1.06 & 27.98 & 1.41 \\
        \rowcolor{configgray}
        18 & @a2ui/\allowbreak web\_core & 2.73 & 0.22 & 3.18 & 0.68 & 38 & @pdfme/\allowbreak common & 1.55 & 0.20 & 3.98 & 0.20 \\
        19 & @pdfme/\allowbreak schemas & 1.78 & 0.79 & 6.39 & 0.73 & 39 & mathlive & 5.48 & 1.52 & 142.45 & 4.49 \\
        \rowcolor{configgray}
        20 & @payloadcms/\allowbreak plugin-\allowbreak ecommerce & 1.46 & 0.22 & 5.73 & 0.70 & 40 & node-\allowbreak forge & 1.57 & 1.25 & 15.52 & 1.91 \\
        \midrule
        \multicolumn{8}{@{}l}{\textbf{Average across all 40 programs}}
        & \textbf{4.35} & \textbf{1.60} & \textbf{45.01} & \textbf{1.94} \\
        \bottomrule
    \end{tabularx}
\end{table*}

\section{Supplementary Evaluation}
\label{s:appendix-supp}

\subsection{Dataset Vulnerability Types}
\label{s:appendix-dataset-types}
\autoref{tab:datasets} gives the full distribution of vulnerability types across the 40 evaluation programs (\autoref{s:eval-setup}).

\begin{table}[t]
    \centering
    \caption{Vulnerability types in the evaluation dataset.}
    \label{tab:datasets}
    \small\rmfamily\upshape
    \setlength{\tabcolsep}{6pt}
    \begin{tabularx}{\columnwidth}
        {@{}>{\raggedright\arraybackslash}Xrr@{}}
        \toprule
        \textbf{Vulnerability type} &
        \textbf{Count} & \textbf{Share} \\
        \midrule
        Cross-site scripting & 12 & 30.0\% \\
        Path traversal & 8 & 20.0\% \\
        Authorization bypass & 5 & 12.5\% \\
        Command injection & 3 & 7.5\% \\
        Prototype pollution & 2 & 5.0\% \\
        SQL injection & 2 & 5.0\% \\
        Code injection & 2 & 5.0\% \\
        Others\textsuperscript{\textdagger} & 6 & 15.0\% \\
        \midrule
        {Total} & {40} & {100.0\%} \\
        \bottomrule
    \end{tabularx}
    \par\smallskip
    \parbox{\columnwidth}{\footnotesize\raggedright
        \textsuperscript{\textdagger}Cryptographic validation,
        SSRF, credential leakage, ReDoS, business logic,
        and markup injection (one each).}
\end{table}

\subsection{Rule Usage per Program}
\label{s:appendix-rule-usage}
\begin{figure}[t]
    \centering
    \includegraphics[width=\columnwidth]{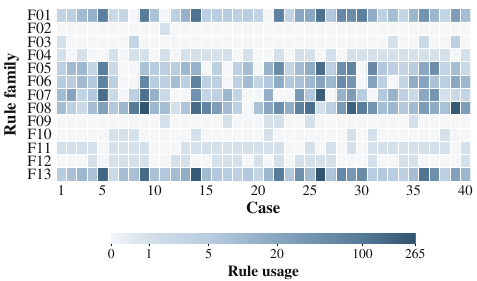}
    \caption{Rule usage per case. Each cell shows how many times a family is used in that case, on a log scale.
    \cc{F04} and \cc{F09}--\cc{F12} are used at most once per case, so their rows stay light even when they appear in many cases.
    }
    \label{fig:heatmap}
\end{figure}

\autoref{fig:heatmap} shows how many times each rule family is applied to each of the 40 evaluation programs (\autoref{s:eval-rules}).

\begin{table}[t]
    \centering
    \caption{Defense effectiveness and overhead of the Python-language extension.}
    \label{tab:python_extension}
    \scriptsize\rmfamily\upshape
    {
    \begin{tabular}{@{}lccccc@{}}
        \toprule
        \textbf{Setting}
        & \textbf{DSR}
        & \textbf{Recall}
        & \textbf{Precision}
        & \textbf{Time}
        & \textbf{Cost} \\
        & \textbf{(\%)}
        & \textbf{(\%)}
        & \textbf{(\%)}
        & \textbf{(min)}
        & \textbf{(\$)} \\
        \midrule
        Unprotected & -- & 70.00 & 46.67 & 31.03 & 6.61 \\
        \sys-Py & 42.86 & 40.00 & 36.36 & 43.62 & 7.25 \\
        \bottomrule
    \end{tabular}%
    }
\end{table}

\subsection{Extending Rules to Python}
\label{s:appendix-python}

To test whether the extracted rules apply beyond JavaScript, we port them to Python.
Following the construction of the JavaScript dataset, we collect ten Python cases from real-world packages, each containing one ground-truth vulnerability.
We randomly sample 60 of the 111 rules, covering all 13 rule families, and manually adapt them to Python.
For each rule, we adapt its matching conditions and operations to the Python AST, preserve indentation, and check the syntax after each application.
We refer to the resulting system as \sys-Py and use Codex with GPT-6.1-sol as the detection agent.

As shown in \autoref{tab:python_extension}, \sys-Py reduces Codex's recall from 70.0\% to 40.0\%, a DSR of 42.9\%.
It also lowers precision from 46.7\% to 36.4\% and increases detection time and cost by 1.41\x and 1.10\x, respectively.
The DSR is lower than on JavaScript (52.0\% for Codex), as \sys-Py is a prototype that ports only part of the rule set by hand.
Our goal here is to show whether the rules apply to other languages rather than to build a complete Python defense.
The result suggests that the rule families capture strategies that are not specific to JavaScript.
The effectiveness can be further improved by porting the remaining rules or by discovering rules directly on Python programs.

\subsection{Distribution of Hidden Vulnerabilities}
\label{s:appendix-overlap}

\begin{figure}[t]
    \centering
    \includegraphics[width=0.56\columnwidth]{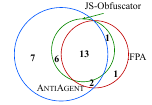}
    \caption{Overlap of vulnerabilities hidden by defense methods.}
    \label{fig:codex_missed_overlap}
\end{figure}

\autoref{fig:codex_missed_overlap} shows which vulnerabilities each method hides from Codex.
A vulnerability is hidden if Codex fails to report it on the transformed program.
In total, 30 vulnerabilities are hidden by at least one method.
\sys hides 28 of them, including seven that neither baseline hides.
Only two are hidden by a baseline but not by \sys.
\sys thus hides nearly all vulnerabilities that the baselines hide, as well as many that they cannot.
Note that these sets may include vulnerabilities that Codex already misses on the unprotected programs (15 of 40), so the counts are higher than those implied by DSR.

We manually compared Codex's traces on the two programs that only a baseline hides, \cc{whistle} and \cc{@roomi-fields/notebooklm-mcp} (Cases~11 and~21 in \autoref{tab:application_cost}) to understand the difference.
In \cc{whistle}, both baselines divert Codex from the file-read vulnerability.
Under JS-Obfuscator, Codex spends many steps decoding about 170 renamed and encoded files, then investigates unrelated authorization and certificate-export code.
Under FPA, it focuses on an undefined helper and startup errors introduced by the rewrite.
Under \sys, the security-relevant flow remains short, so Codex decodes the few relevant names and follows the requested filename into the file-reading code.
It finds that the temporary-file restriction is skipped when the owner's credentials are configured.
In \cc{@roomi-fields/notebooklm-mcp}, \sys checks a modified directory path but still creates the directory with the original path, and Codex finds the missing restriction at that call.
Under FPA, Codex reads the file-writing code but overlooks how the filename prefix allows files to be written outside the intended location.
In both cases, the baselines hide the vulnerability only because they divert Codex to other issues.

\subsection{Baseline PoC Failures}
\label{s:appendix-poc-failures}
Under \sys, all 40 PoCs still trigger, whereas 37 trigger under JS-Obfuscator and 33 under FPA (\autoref{tab:transformed_program_quality}).
To understand why the PoCs fail under the baselines, we manually analyze the ten failed cases.
For JS-Obfuscator, two transformed programs stop during module loading and one times out, so the PoC never reaches the vulnerable code.
For FPA, five programs fail to run due to four loading errors and one unresolved dependency.
In the other two FPA cases, path-handling changes or incorrect escaping alter the program's behavior so that the original PoC conditions no longer hold, although the unsafe flow remains.

\end{document}